\documentclass[twocolumn]{aastex631}
\usepackage{subcaption}
\usepackage{float}

\begin{document}
\newcommand{\revone}[1]{{\color{red} #1}}
\newcommand{\revtwo}[1]{{\color{blue} #1}}

\title{Earth as an Exoplanet: Investigating the effects of cloud variability on the direct-imaging of atmospheres}

\author[0009-0005-5118-4174]{Soumil Kelkar}
\affiliation{Indian Institute of Science Education and Research Pune,  Dr Homi Bhabha Road, Pashan, Pune, Maharashtra - 411008, India}
\affiliation{ NASA Goddard Space Flight Center, 8800 Greenbelt Road, Greenbelt, MD 20771, USA}
\affiliation{Center for Research and Exploration in Space Science and Technology, NASA/GSFC, Greenbelt, MD 20771, USA}


\author{Prabal Saxena}
\affiliation{ NASA Goddard Space Flight Center, 8800 Greenbelt Road, Greenbelt, MD 20771, USA}

\author[0000-0002-5893-2471]{Ravi Kopparapu}
\affiliation{ NASA Goddard Space Flight Center, 8800 Greenbelt Road, Greenbelt, MD 20771, USA}

\author[0000-0002-3932-3603]{Joy Monteiro}
\affiliation{Departments of Earth and Climate Science and Data Science, Indian Institute of Science Education and Research Pune, Dr Homi Bhabha Road,
Pashan, Pune, Maharashtra - 411008, India}



\begin{abstract}


A planet’s spectrum is dynamic and only represents a time-dependent snapshot of its properties. Changing atmospheric conditions due to climate and weather patterns, particularly variation in cloud cover, can significantly affect the spectrum in ways that complicate the understanding of a planet's baseline atmospheric properties. Variable cloud cover and cloud properties affect the detectability of atmospheric constituents, and also greatly influence the radiative transfer that determines a planet's spectrum. This has considerable implications for direct imaging observations of potentially habitable exoplanets and thus it is critical to study and characterize the effects of clouds on their spectra. Clouds have been extensively modeled before and their effects have been incorporated across climate frameworks spanning a spectrum of complexity. Given the challenges associated with modeling clouds, we adopt a novel approach in this work to study the effects of clouds by using real-time cloud data from Earth observations. Treating Earth as an exoplanet and using detailed observations from the MERRA-2 data collection, we quantify the effects of cloud variability on the spectrum as well as on the detectability of atmospheric constituents, specifically biomarkers like $\text{O}_2$, $\text{O}_3$ and $\text{H}_2\text{O}$. The coverage and vertical position of clouds significantly affects the SNRs of these gases and subsequently their detectability in exo-Earth atmospheres. Moreover, we show that variations in the amount of cloud cover will potentially confound efforts to retrieve a stable baseline atmosphere for a planet. This work has important applications to future direct-imaging missions like the Habitable Worlds Observatory (HWO).
 
 

\end{abstract}



\section{Introduction} \label{sec:intro}

Amongst all the 5700 exoplanets detected so far\footnote{\url{https://exoplanetarchive.ipac.caltech.edu/}}, there is a significant dearth of exoplanets with properties similar to the inner rocky worlds of our Solar System, including Earth (see figure 1 in \cite{Gaudi_2021}). This deficiency is mainly in part due to the higher sensitivity of current detection methods toward massive planets on short orbital periods. To address this issue, the `Pathways to Discovery in Astronomy $\&$ Astrophysics for the 2020s' report \citep{decadal_survey} recommended an ultraviolet/optical/infrared space observatory to find and characterize Earth-like planets around nearby stars, specifically the ones which might be potentially habitable. Based on this recommendation, NASA is laying the groundwork for its next flagship mission, currently referred to as the Habitable Worlds Observatory (HWO) which draws from previous mission concepts like the LUVOIR \citep{luvoir2019luvoir} and Habex \citep{gaudi2020habitable}. A significant part of the science pre-cursor work for these mission concepts is devoted to simulating reflected light spectra of planets, computing exposure times, and quantifying the detectability of important atmospheric constituents and biosignatures. However, the surface of the planet and/or its atmospheric composition may not be homogeneous, and thus a planet's spectrum is not static but rather dynamic. Moreover, natural variation in cloud coverage due to weather and climate patterns can also significantly affect the characterization of exoplanet atmospheres.

Clouds are ubiquitous on Earth and other solar system planets, and observations strongly suggest their presence on exoplanets as well (\cite{line2013near,sing2015hst,kreidberg2014clouds}). A cloud is defined as the condensate that forms when the vapor pressure of an atmospheric constituent exceeds its saturation vapor pressure (\cite{marley2013clouds}). In this way, cloud formation provides a sink for an atmospheric constituent. Earth is dominated by clouds composed mostly of water vapor, however, there is a huge diversity in the composition of clouds on planets in our Solar System. This includes sulphuric acid clouds on Venus (\cite{hansen1974interpretation}), $\text{CO}_2$ ice clouds on Mars (\cite{montmessin2007hyperspectral}), and ammonia clouds on Jupiter (\cite{brooke1998models}), to name a few. Given the diversity in the estimated planetary and atmospheric compositions of exoplanets detected so far, we would also expect their atmospheres to host clouds and other aerosols of varied compositions. A haze is referred to as any condensate produced by photochemistry or other non-equilibrium chemical processes (\cite{marley2013clouds}), however, the term is often used interchangeably with cloud. A general framework of clouds in the atmospheres of Solar system planets is given in \cite{sanchez2004clouds} and a comprehensive summary of clouds and cloud formation in exoplanet atmospheres is given in \cite{marley2013clouds, helling2019exoplanet, gao2021aerosols}.

Clouds affect a planet's radiation budget in three important ways - (i) by reflecting/scattering shortwave stellar radiation back to space, (ii) by absorbing and re-emitting longwave thermal radiation emitted by the surface, and (iii) also by emitting their own thermal radiation. Clouds can complicate the radiative transfer processes in an exoplanet's atmosphere and significantly impact the planet's spectrum. Clouds tend to have high albedos, especially in the visible wavelength band, and thus any planet with a cloudy atmosphere will reflect much more light than a cloudless planet, enhancing the continuum in the reflected light spectrum. Clouds improve the detectability of atmospheric constituents in the visible band, especially those present in significant amounts above the cloud layer, by increasing the albedo and subsequently boosting their absorption signals (\cite{kawashima2019theoretical,wang2017baseline}, section 3.2 of this work). The effects of clouds on the reflected light spectra were modelled by \cite{marley1999reflected,Sudarsky_2003,Cahoy_2010} for giant exoplanets and by \cite{Kitzmann_2011b} for Earth-like exoplanets. Since clouds enhance a planet's albedo, they may impact whether a planet is visible or not via direct imaging. The impact of clouds on the planet's albedo is explored in \cite{Kitzmann_2011a}. Clouds are a significant contributor to the greenhouse effect as they absorb and re-emit outgoing longwave radiation emitted by the surface. Thus, the presence of clouds has a profound impact on the thermal emission spectrum of a planet. Due to the greenhouse effect, clouds dampen the spectral features of certain gases in the infrared. The effects of clouds on the thermal emission spectra of planets have been discussed in \cite{tinetti2006detectability, Hearty_2009, Kitzmann_2011a, vasquez2013infrared}. Furthermore, clouds also affect spectral signatures originating from the surface, for example, the vegetation red edge signal characteristic of surface vegetation, as shown in models (\cite{tinetti2006b,montanes2006vegetation}) and observations of Earth (\cite{Arnold_2002,hamdani2006biomarkers}).

Clouds have been extensively modelled with different models spanning a spectrum of complexity. The simplest of cloud models are parametrized models (\cite{gao2021aerosols} and references therein) which are typically used for retrieval studies. These models do not include any treatment of the physical processes associated with cloud formation and composition but simply use a set of parameters to estimate the first order effects of clouds. More complex models, on the other hand, involve cloud microphysics and they typically assume a functional form for the cloud particle size distribution. These cloud models are incorporated into general climate model frameworks ranging from 1D radiative-convective models to 3D general circulation models (GCMs). 1D climate models, by virtue of their definition, cannot estimate differences in horizontal cloud coverage and thus cannot model atmospheres with patchy clouds. Naturally, they also neglect significant multi-dimensional circulation phenomena which impact cloud properties. 3D GCMs can accurately model the spatial extent of clouds on a planetary scale, however they can sometimes be too computationally expensive to run, especially if complex cloud microphysical processes are involved. Uncertainties in quantifying cloud feedbacks and difficulties in resolving clouds on large model grids (\cite{Kofman_2024}) signify that there are still challenges involved in the accurate representation of clouds in 3D GCMs. There are additional challenges to modelling clouds on terrestrial exoplanets due to a lack of observational constraints on the type and composition of their atmospheres. Uncertainties in observed atmospheric compositions and theoretical predictions make it difficult to accurately estimate the composition and spatial extent of clouds. Nevertheless, clouds have been modelled on terrestrial exoplanets using a variety of approaches. The simplest models modify the surface albedo of the planet to mimic the effects of clouds (\cite{kasting1993habitable,segura2003ozone,segura2005biosignatures,grenfell2007response}). In this approach, there are no physical assumptions made about the properties of clouds and it does not impact radiation in the same way as real clouds, hence, these models cannot be used to study the impact of clouds on exoplanet spectra. In more complex models, the properties of clouds and the atmosphere itself are assumed or are constrained to some extent via observations. This approach has been used to model clouds on Earth-like exoplanets using Earth as a reference (\cite{kaltenegger2007spectral,kitzmann2010clouds}). For a more comprehensive description of the the intricacies of cloud modelling and the different types of cloud models present in literature, refer to \cite{marley2013clouds} and \cite{gao2021aerosols}.       

Given the challenges associated with modelling clouds, one strategy for an accurate examination of the impact of clouds and cloud variability on terrestrial Earth-like planets involves using real-time cloud data from Earth observations. Several Earth-observing satellites collect real-time data crucial for climate monitoring and weather forecasting. This data is assimilated into data products which can be extracted to obtain atmospheric constraints and real-time values of atmospheric temperature, mixing ratios of gases and more importantly for this study, cloud properties. Earth's patchy cloud cover is unique amongst its terrestrial neighbours. Real time data from satellites can be leveraged to study the temporal and spatial extent, composition and albedo of clouds. Data gathered over different portions of the globe can be integrated to produce full-disc observations and global maps of cloud cover. Such integrated disc observations are essential to quantify the impact of clouds on the planetary spectrum, as future direct imaging missions will only have a full integrated disc view of exoplanets. Moreover, such globally integrated data can be utilized to assess the impact of cloud variability across different temporal and spatial scales on planetary spectra. Real time data has been extensively used to study cloud properties and quantify trends in cloud coverage over different scales (\cite{king2013spatial,wylie2005trends,stowe1991global,wu2011global,rossow1990global}). Earth's cloud cover shows significant variation across different spatial and temporal scales. Clouds are non-uniformly distributed across different latitudes. The cloud cover over oceans is significantly higher than land cloud cover. Cloud cover variations over diurnal and seasonal scales are also well documented (\cite{king2013spatial}). Such significant cloud variability influences full disc observations and subsequently impacts the planet's spectrum.   

The studies described above which have modeled the effects of clouds on reflection and emission spectra, planetary albedo have relied on parameterized cloud models. In this work, we adopt a novel approach which involves using real-time empirical cloud data from Earth observations to study cloud effects. This allows us to incorporate realistic cloud distributions which are more accurate than the distributions simulated by cloud models, as well as analyze the effects of naturally varying cloud cover. Our results computed using empirical 3D cloud distributions complement previous results obtained using cloud models.  For example,  \cite{Kitzmann_2011a} and \cite{kawashima2019theoretical} use a 1D climate model coupled with a parametric cloud model to quantify the effects of different amounts of low-level and high-level cloud cover on the reflection spectra and we complement their results (in section 3.3).    

In the context of the future direct-imaging missions whose main goal is to find and characterize nearby Earth-like exoplanets, an investigation into the impacts of clouds on the characterization of exo-Earth atmospheres stands as an important science pre-cursor. In this study, we treat Earth as an exoplanet, and utilize real time data from the Modern-Era Retrospective analysis for Research and Applications, Version 2 (MERRA-2) (\cite{gelaro2017modern}) dataset to 
conduct a detailed analysis of the effects of cloud variability on its spectrum.   
We also conduct a comparison between all the future direct-imaging mission concepts and evaluate their performance in characterizing different exo-Earth atmospheres with varying cloud cover.

\section{Methods} \label{sec:methods}

In this study, we use real time remote sensing Earth data to construct an accurate 3D model of an Earth twin or an exo-Earth. 
The Modern-Era Retrospective analysis for Research and Applications, Version 2 (MERRA-2) (\cite{gelaro2017modern}) is a NASA atmospheric reanalysis produced with the Goddard Earth Observing System, Version 5 (GEOS-5) data assimilation system. Reanalysis is a process where an already existing data assimilation system is used to reprocess meteorological observations. MERRA-2's continuous data record spans from 1980 to the present and contains either instantaneous or time-averaged data products. MERRA-2 relies on an underlying forecast model to combine raw data from disparate observations in a physically consistent manner, enabling production of gridded datasets for a broad range of variables (\cite{gelaro2017modern}).   
MERRA-2 has several data collections, each corresponding to different temporal resolutions, surface and atmospheric parameters, aerosol properties etc. Specifically, we use the M2I3NVASM data collection (\cite{merra2_data_citation}) which is an instantaneous 3-dimensional gridded data collection that consists of assimilations of meteorological parameters. Each data file is a snapshot of atmospheric properties, captured every 3 hours. We extract all the relevant atmospheric parameters such as temperature, wind speeds, and mixing ratios of ozone, water vapor, liquid water, and water-ice clouds (constituents which are not well-mixed in the atmosphere). The data has a temporal resolution of 3 hours and a horizontal spatial resolution of $0.5^\circ$ latitude $\times$ $0.625^\circ$ longitude. The atmosphere is divided into 72 model levels from the surface to 0.01hPa. Thus, this data collection allows us to examine cloud variability at different temporal scales and subsequently construct a dynamic 3D exo-Earth model with cloudy patches which exhibit significant variations over time. Spatial variations over latitudinal scales are averaged over since we are considering a planet at quadrature orbiting in an edge-on orbit. These variations would be much more relevant when considering planets with a non-negligible system inclination relative to the observer (e.g. oriented in a face-on or a near face-on configuration). Hence, when considering the impacts of cloud variability, we are only concerned with temporal variations in a globally-averaged cloud cover .


To construct the ground map, we use data from the MODerate resolution Imaging Spectroradiometer (MODIS) (\cite{modis_data}) instrument aboard the Terra and Aqua satellites, which takes high-resolution maps of the Earth's surface. This allows us to accurately replicate Earth's ocean and landmass distributions in our exo-Earth model. MODIS data has a resolution of $0.05^\circ$ latitude $\times$ $0.05^\circ$ longitude. We assume that the surface is static, in order to eliminate any changes on the surface (for eg. change in albedo due to varying snow cover or seasons) from impacting the spectra. In this way, we can attribute any change in spectra to be caused by variations in atmospheric conditions, specifically cloud variability.


\begin{figure*}[!ht]
  \centering

  \begin{subfigure}{0.45\textwidth}
    \centering
    \includegraphics[width=\linewidth]{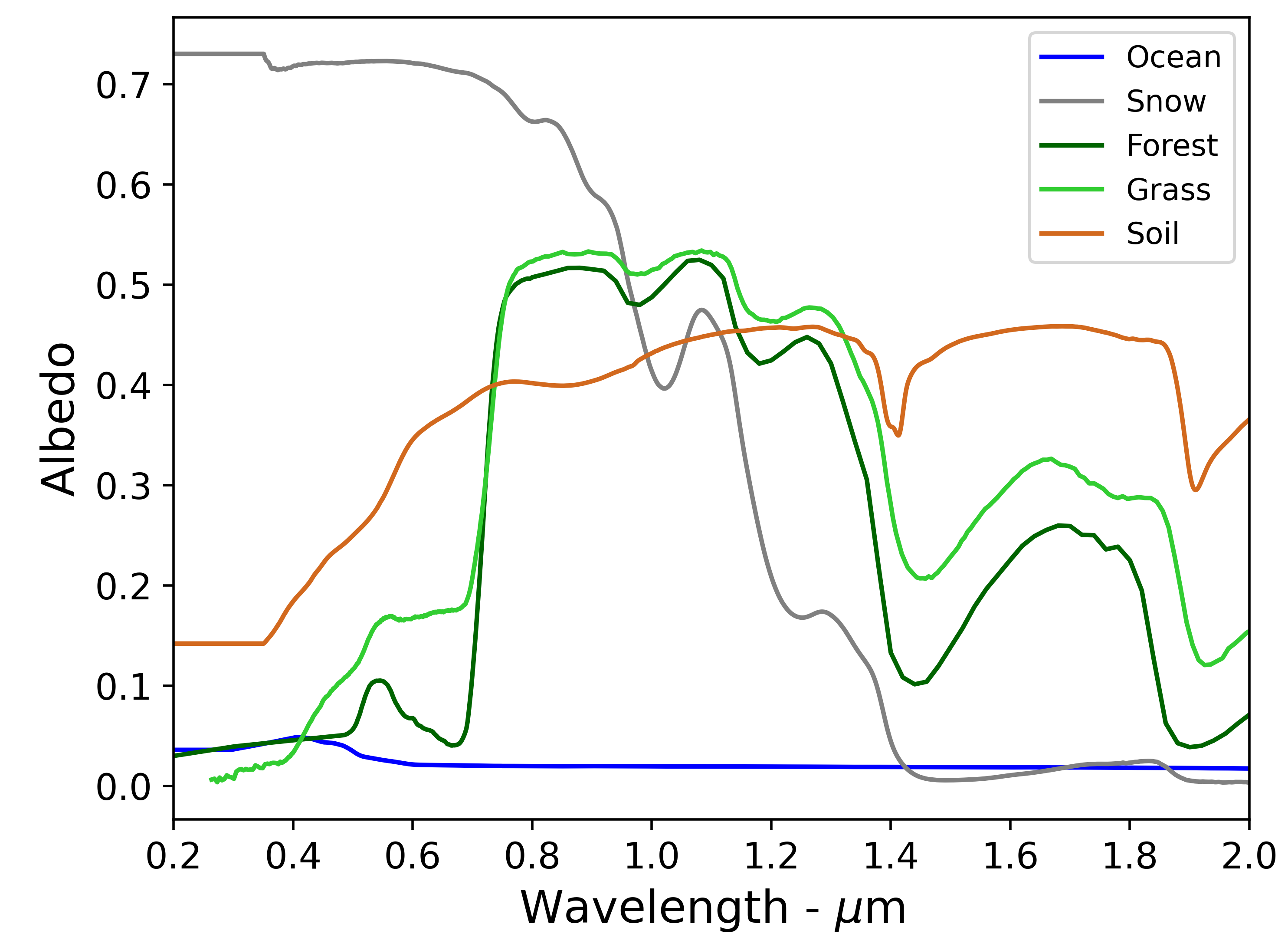}
    \label{fig:surf_albedo}
  \end{subfigure}%
  \hfill
  \begin{subfigure}{0.45\textwidth}
    \centering
    \includegraphics[width=\linewidth]{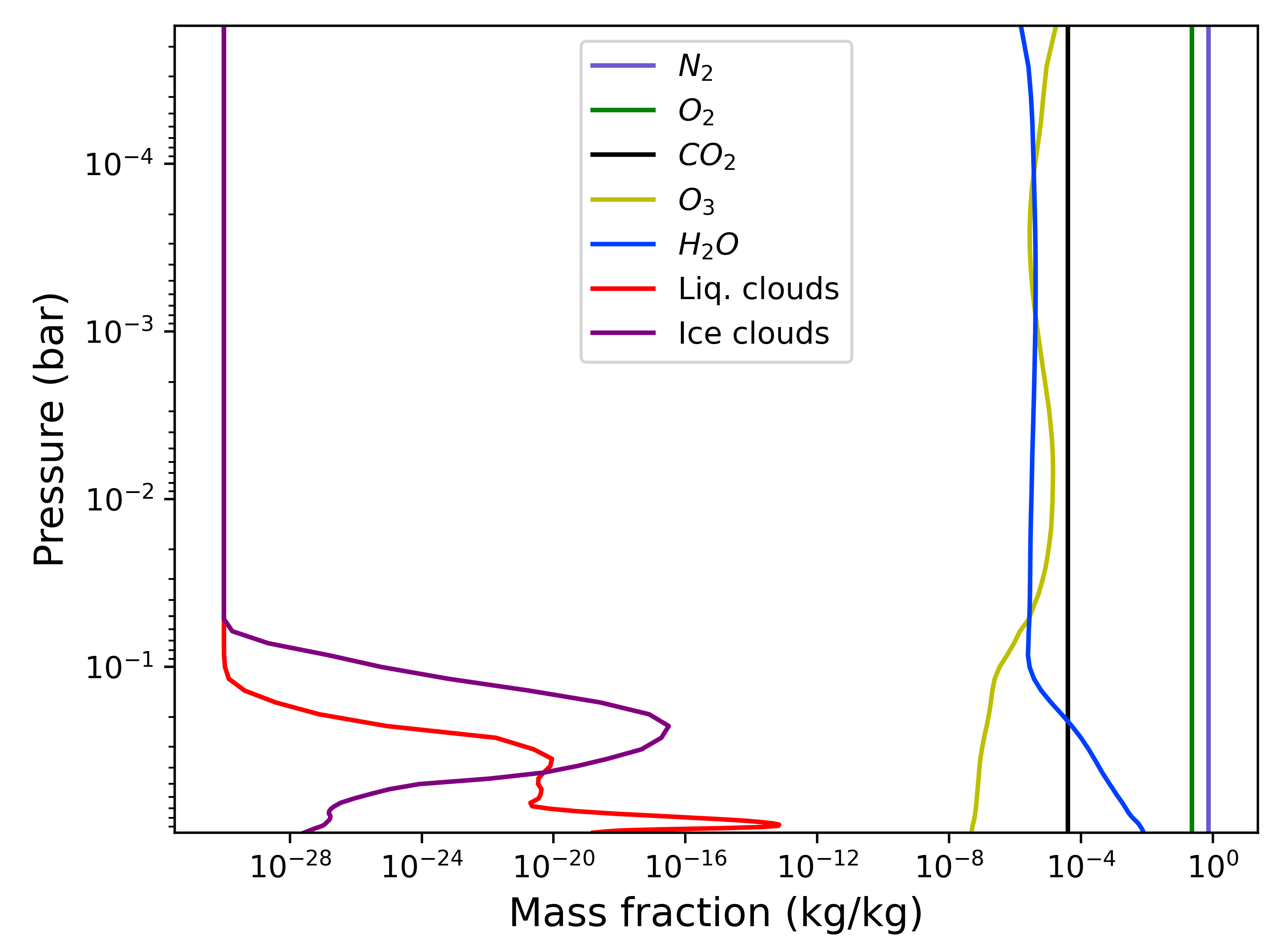}
    \label{fig:vert_prof}
  \end{subfigure}

\caption{(a) The wavelength-dependent surface albedos of the five surface types - ocean, snow, grass, soil, and forest. The albedo values have been taken from the USGS spectral library (\cite{kokaly2017usgs}). The significant increase in the albedos of forest and grass at roughly 0.7$\mu$m corresponds to the characteristic vegetation red edge signal which indicates the presence of surface vegetation. (b) The vertical abundance profiles for the dominant gases and aerosols present in the atmosphere. The abundance of each constituent is denoted by its disk-integrated mass fraction in each layer of the atmosphere. These are modern Earth values derived from the MERRA-2 data corresponding to 20th March 2019 00:00 UTC.}
  \label{fig:atm_profile}
\end{figure*}


Given our exo-Earth model, we use the Planetary Spectrum Generator (PSG) (\cite{villanueva2018planetary}) to simulate observations of this planet. PSG is a radiative transfer model suite that combines state-of-the-art radiative transfer codes, spectroscopic and planetary databases to accurately synthesize and retrieve planetary spectra for any given planetary system. It can simulate observations for a broad range of wavelengths from any given observatory or mission concept. We specifically use the Global Emission Spectra (GlobES) module of PSG which ingests our 3D exo-Earth model and accurately simulates the spectrum given a range of observational and instrumental parameters. The different aspects of the calculations done will be explained here briefly, but to find out more about using PSG, please refer to the web interface (\url{https://psg.gsfc.nasa.gov}), the original paper describing the tool (\cite{villanueva2018planetary}), a recent paper describing use of PSG to simulate directly imaged exoplanet spectra (\cite{Saxena_2021}) or the PSG documentation handbook (\cite{psg_handbook}). 

GlobES can ingest GCM files which are converted from the typical netCDF data format. Based on the python codes supplied on the PSG Github page (\url{https://github.com/nasapsg/globes}), we develop an efficient pipeline that retrieves data from the MERRA-2 online data repository (\url{https://goldsmr5.gesdisc.eosdis.nasa.gov/opendap/hyrax/MERRA2/M2I3NVASM.5.12.4/}), constructs an accurate 3d model of an exo-Earth from this empirical data, converts the model into a GCM file which can be uploaded to GlobES, and finally retrieves the simulated spectrum. In addition to atmospheric data, the GCM file also includes information about the orbital configuration of the planet, the geometry of the planetary system with respect to the observer, the specific details of the instrument used to observe the system, and a noise model that incorporates different sources of noise. The entire list of parameters which can be specified in the GCM file can be found on the PSG web interface (\url{https://psg.gsfc.nasa.gov/helpapi.php#parameters}). 

We assume a simplified fiducial exo-Earth that is situated 10 parsecs away from the observer, orbiting a Sun-like G star in a circular orbit, such that the orbit is oriented in an edge-on fashion. Given an observing geometry, PSG computes the set of incidence and emission angles using a set of sampling algorithms. These angles ultimately inform the radiative transfer codes which compute the amount of light reflected by the planet at each wavelength (see Chapter 2 of the PSG handbook \cite{psg_handbook}). The atmosphere is assumed to be in hydrostatic equilibrium and is composed of $\text{N}_2$, $\text{O}_2$, $\text{CO}_2$, $\text{CH}_4$, $\text{N}_2\text{O}$, $\text{CO}$, $\text{H}_2\text{O}$ and $\text{O}_3$ along with liquid water and water-ice clouds. The mean vertical abundance profiles of the dominant gases and aerosols are given in figure \ref{fig:atm_profile}. $\text{N}_2$ and $\text{O}_2$, $\text{CO}_2$, $\text{CH}_4$, $\text{N}_2\text{O}$ and $\text{CO}$ are equally distributed throughout the atmosphere with abundances of $78\%$ and $20.9\%$, 400 ppmv, 1.6 ppmv, 1.3 ppmv, and 0.1 ppmv respectively. $\text{H}_2\text{O}$ and $\text{O}_3$ are not equally distributed and their profiles, as shown in figure \ref{fig:atm_profile} are derived from the MERRA-2 data. Similarly, the vertical and horizontal distribution of clouds is derived from the empirical data. In our model, the liquid water and water-ice cloud particles are assumed to have a constant size of $5\mu\text{m}$ and $100\mu\text{m}$ respectively. 

For low and mid-resolution simulations, PSG uses correlated k-tables to model the absorption of gases and these are based on the HITRAN/HITEMP database (\cite{gordon2022hitran2020,rothman2010hitemp}). These correlated k-opacity tables are precomputed in PSG to make the radiative transfer calculations more efficient. Apart from modeling absorption of gases, we model different atmospheric processes including Rayleigh scattering, refraction, collision-induced absorption, and UV absorption. Rayleigh scattering is computed in PSG following the methodology given in \cite{sneep2005direct}. PSG uses the refraction indices from \url{https://refractiveindex.info} to model refraction in the atmosphere, CIA datasets from HITRAN (\cite{gordon2022hitran2020} and references therein) to model CIA and the MPI-Mainz UV/VIS Spectral Atlas (\cite{keller2013mpi}) and other UV databases to model UV absorption. 

 We assume that the surface is Lambertian, and thus it scatters light isotropically in all directions. For simplicity, we divide the ground coverage into 5 surface types - ocean, snow, grass, soil, and forest, whose wavelength-dependent albedos are shown in figure \ref{fig:atm_profile}, where the albedo values are taken from the USGS spectral library (\cite{kokaly2017usgs}). To compute the disk-integrated flux reflected from the surface, PSG uses the parameters given to describe the observing geometry and the direction of incidence and reflected fluxes. The incidence and reflection angles, along with the albedo of the surface and other parameters are captured in the bi-directional reflectance distribution function (BRDF) that describes how light from a source is reflected/scattered by an opaque surface. The surface of the planet is not homogeneous and each of the 5 surface types has different optical properties. Since we cannot resolve individual surface types, we employ the areal linear mixing model to calculate the effective optical properties of the aggregate. This model assumes that the entire disc is composed of smaller unresolved patches made up of different surface types and that each patch is homogeneous and can be treated separately. The properties of the entire disc like the geometric albedo are simply a linear sum of the individual optical properties of each patch weighted by the corresponding fractional area occupied in the disc. 

To perform radiative transfer calculations and accurately simulate spectra, PSG utilizes the Planetary and Universal Model of Atmospheric Scattering (PUMAS) which integrates radiative transfer codes, spectroscopic parameterizations and correlated-k tables to compute spectra (\cite{villanueva2015strong,wolff2009wavelength,edwards_genln2}). Our 3D exo-Earth model is ingested into the GlobES module of PSG where the 3D atmospheric data is mapped onto a 2D observational grid which is fed to the radiative transfer code. The whole disc is sampled according to the bin size which is specified in the GCM file. PSG performs the radiative transfer calculations across the whole observable disc and the disc-integrated spectrum is thus calculated as the linear sum of individual spectra weighted by the projected area of each bin. To consider the effects of Rayleigh scattering and scattering by clouds and other aerosols, the radiative transfer calculations should involve multiple scattering processes. For this purpose, PSG uses the DISORT package which was specifically developed to efficiently solve the multiple scattering problem (\cite{stamnes1988numerically,stamnes2000disort,buras2011new}). The scattering problem is solved by using a set of numerical approximations which discretize the differential radiative transfer equation. The number and the size of the discrete equations describing the scattering function can be varied to determine the accuracy of the solution. The number of equations is encoded in `NMAX' and the size of each equation is given by `LMAX'. For our simulations, we choose NMAX=1 and LMAX=66. For more details about the radiative transfer calculations and multiple scattering analysis performed by PSG, refer to chapters 4 and 5 of the handbook \citep{psg_handbook}. 

\subsection{Computing Signal-to-Noise Ratios (SNRs)}

To calculate the SNRs, we adopt a method that has been well-established in the literature (\cite{Kopparapu_2021_no2,checlair2021probing}). To compute the SNR for the detection of any atmospheric constituent, we simulate spectra for two different atmospheric models - (i) the default model which includes all atmospheric gases, and (ii) the missing gas model which includes all gases except one. We take a difference of the spectra simulated using these two models (this is the signal) and divide it by the noise simulated by PSG (see Chapter 8 of the PSG handbook \cite{psg_handbook}) to get the SNR at each wavelength. The net SNR can be calculated as a square root of the sum of squares of individual SNRs at each wavelength. 

\[ \text{netSNR} = \sqrt{\sum_{\text{i}}(\text{SNR})_{\text{i}}^2} \]


\section{Results} \label{sec:results}

\subsection{Spectra of a rotating and revolving exo-Earth}

We simulate observations of the planet at quadrature, where it is at the maximum angular separation from the star, and well outside the inner working angle of the coronagraph. We choose a random day (1st July 2000) from MERRA-2's historical record and construct our atmospheric model based on empirical data obtained on that day. We also looked at other days and based on the cloud distribution on any given day, we observed a similar trend in the spectra, and hence, we only present results for one case.

Figure \ref{fig:rot_exoearth_spec} shows the simulated spectra of a rotating planet where each rotational configuration is labeled by a rotational phase. Each rotational phase is separated by 3 hours to match the temporal resolution of the MERRA-2 atmospheric data. At each phase, we update our atmospheric model with new empirical data to ensure that our model accurately replicates the dynamic atmospheric conditions prevalent on Earth. Figure \ref{fig:rot_exoearth_spec} shows the simulated reflected light spectra in three wavelength bands - ultraviolet (UV), visible (VIS), and near-infrared (NIR) for the different rotational phases labeled in the top panel. The range of the wavelength bands is adopted according to the recommendations of the LUVOIR mission report (\cite{luvoir2019luvoir}). 
The cloudy planet's spectrum shows significant variation over a short span of 24 hrs and this dynamic nature of the spectrum is primarily attributed to variability in cloud cover over different parts of the disc. This is similar to what has been observed on Earth with DSCOVR \citep{gu2021earth} and in other exo-Earth studies ({\cite{Kofman_2024}). This variation is highly apparent in between 0.3 to 0.7 $\mu$m and 1.5 to 1.8 $\mu$m. The spectra of a cloudy planet show at max a $\sim 75\%$ variation in reflected flux at 0.5 $\mu$m for different rotational phases. In contrast, the cloud-free spectra only vary by $\sim 15 \%$ at the same wavelength, due to the changing surface albedo.Extensive cloud cover over the eastern Pacific Ocean (phase 270) and parts of south and south-east Asia (phase 135), significantly enhance the reflected light continuum at these phases. In contrast, the dearth of clouds over Africa (phase 45) and the western Atlantic Ocean (phase 0) lead to a much lower reflectance.  In the absence of clouds, the reflected light continuum shifts to a lower value as the surface has a lower albedo than the clouds. Additionally, the spectra show less variation, attributed now only to the changing surface albedo. 

\begin{figure*}
    \centering
    \includegraphics[width=0.95\textwidth]{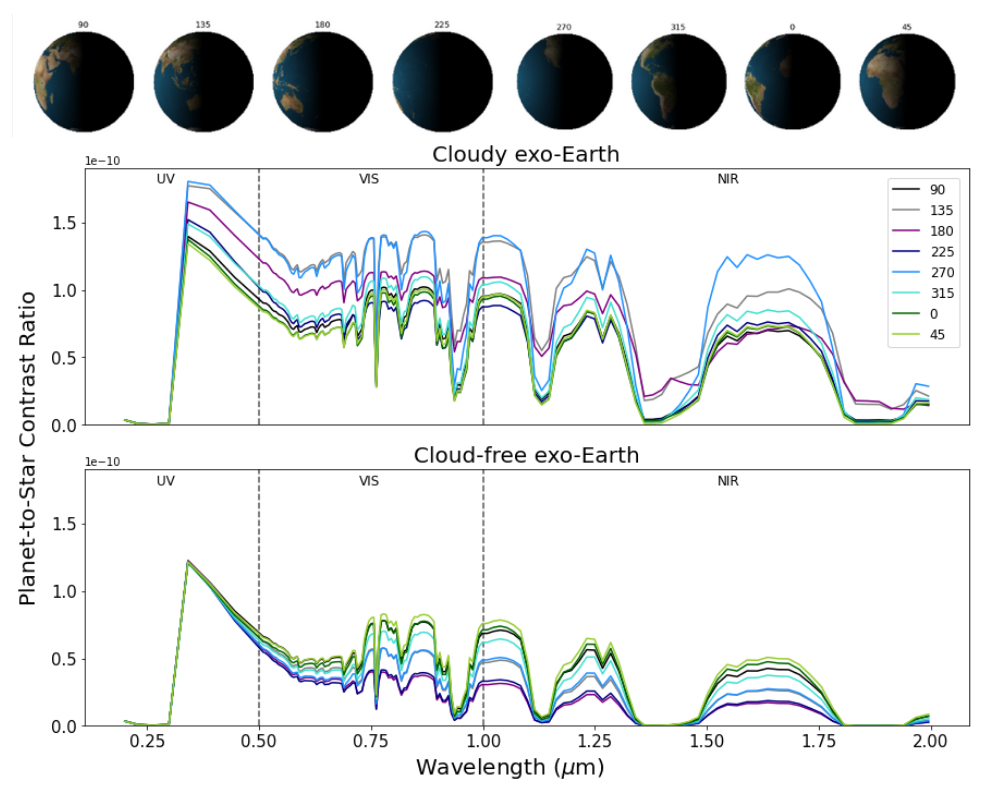}
    \caption{The top panel shows the different rotational configurations of the exo-Earth labeled by a rotational phase. These images only depict how the portion of the surface visible to the observer changes with rotation and don't imply that the planet is cloudless. The plots show the reflected light spectra for the different rotational phases as simulated by PSG for a cloudy and a cloud-free exo-Earth respectively. The entire wavelength bandpass is seperated into three regions - (i) ultraviolet (UV) (0.2-0.5 $\mu$m), (ii) visible (VIS) (0.5-1 $\mu$m) and (iii) near-infrared (NIR) (1-2 $\mu$m). The simulated error bars have been removed for the sake of visual clarity. The cloudy planet shows a higher continuum in the reflected light which is attributed to the enhanced albedo in the presence of clouds. The spectrum of a cloudy planet also shows a higher degree of variation between rotational phases due to variable cloud patchiness.}
    \label{fig:rot_exoearth_spec}
\end{figure*}

We also compare the reflected light spectra of both a cloudy and a clear planet at three different orbital phases - quadrature (Q) and $\text{Q} \pm 30^\circ$, ensuring that the planet is outside the inner working angle of the coronagraph. For these simulations, the exposure time is set to 24 hours. Since the MERRA-2 data collection has a temporal resolution of 3 hours, we combine 8 data files corresponding to a single day (1st July 2000), and take an average of the atmosphere and surface data to construct our exo-Earth model. This is done to replicate a real 24-hr observation, in which, the instrument would receive light reflected by different parts of the planet's disc as it rotates. The collected photon data would represent an integrated disc `averaged over' each rotational configuration. Considering both a cloudy and a clear planet, the spectra show a $\sim 30-100\%$ variation in spectra at $0.5\mu$m between subsequent orbital phases. This is comparable to the variations seen in the spectra of a rotating cloudy planet. The spectra show a lower percentage of variation at higher wavelengths (for eg. $\sim 25-40\%$ variation at $1\mu$m) which implies that the scattering and reflection by clouds in the visible band enhance the changes in spectra at different orbital phases.


\subsection{The impact of clouds on SNRs of atmospheric gases}

\begin{figure*}
    \centering
    \includegraphics[width=0.95\textwidth]{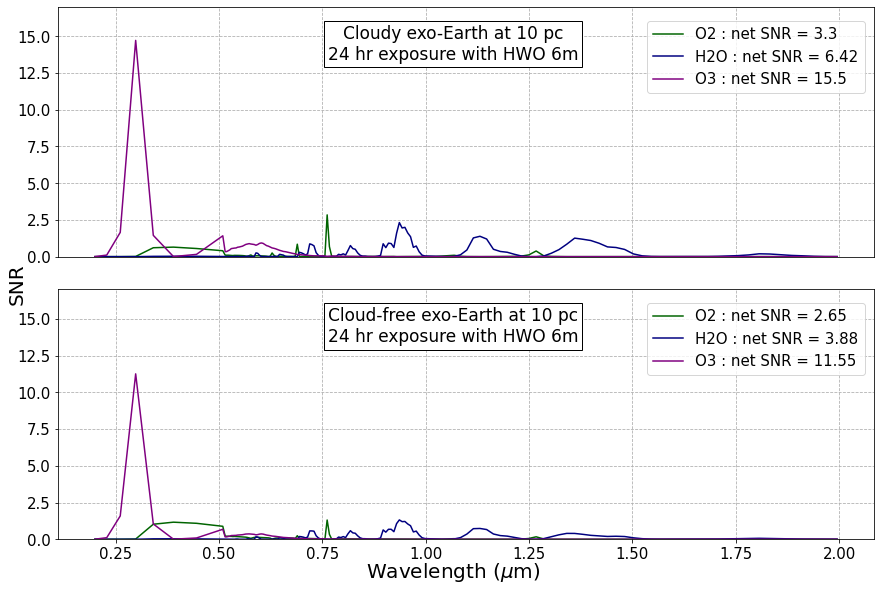}
    \caption{The wavelength-dependent SNRs of the three gases, $\text{O}_2$, $\text{O}_3$ and $\text{H}_2\text{O}$. The top panel shows the results for a cloudy exo-Earth while the bottom panel depicts the same for a cloud-free exo-Earth. The simulations are performed in PSG using a 6m HWO template and a suitable noise model assuming that the exo-Earth is situated 10 pc away and the exposure time is set to 24 hrs. The corresponding net SNRs for each gas across the entire wavelength region is given in the legend. Clouds enhance the SNRs of all the three gases and hence improve their detectability in an exo-Earth atmosphere. The increase in SNR for each gas is dependent on its vertical distribution with respect to the distribution and opacity of patchy clouds.}
    \label{fig:exo_Earth_snr_all_gases}
\end{figure*}

The previous section shows that clouds have a significant impact on the reflected light spectrum of the integrated disc as well as on the absorption signals of different gases. By extension, we explore their impact on the Signal-to-Noise Ratios (SNRs) of these gases in this section. We restrict our analysis to $\text{O}_2$, $\text{O}_3$ and $\text{H}_2\text{O}$ because (i) they have strong absorption lines in the considered wavelength bands and moreover, (ii) they are important bio-markers. Figure \ref{fig:exo_Earth_snr_all_gases} shows the wavelength-dependent SNRs of these three gases for a cloudy and cloud-free exo-Earth, as computed assuming a 6m HWO telescope PSG template with a suitable noise model and a 24 hr exposure. The instrumental parameters of this template are given in table \ref{table:mis_con_par} and figure \ref{fig:thrput}. 
Clouds enhance the SNRs of all the three gases across the considered wavelength range. Since the same noise model is employed for the cloudy and the cloud-free case, the subsequent increase in SNRs is due to an increase in the strength of absorption signals in the presence of clouds which is attributed to their greater (than the surface) albedo. Since its difficult to probe the atmosphere below a cloud layer, for clouds to boost the absorption signals of these gases, they should be present in significant quantities above the cloud layer. Hence, the vertical abundance of each gas with respect to the clouds also influences the SNR values.$\text{O}_3$ is abundant in the stratosphere, much higher than a large fraction of the clouds, while $\text{O}_2$ is well-mixed throughout the atmosphere and $\text{H}_2\text{O} $ is abundant in the troposphere. Even though $\text{H}_2\text{O}$ abundance reduces with height, a significant fraction of the gas is still present above the layer of liquid clouds (see Figure \ref{fig:atm_profile}). Optically thick clouds in the lower altitude regions of the troposphere have a huge impact on the SNRs of $\text{O}_2$ and $\text{H}_2\text{O}$, enhancing them by almost a factor of 2 (e.g. the $\text{O}_2$ absorption line at $0.76\mu$m). Higher up in the troposphere, the clouds get optically thinner and thus their impact on the SNRs of $\text{O}_3$ is not that significant ($\sim$ 11 for a cloud-free case to $\sim$ 15 for a cloudy case at $0.3\mu$m).}The $\text{O}_2$ signature in the UV band might be a computational artifact caused by imprecise modelling of the Rayleigh tail, but we are still investigating the exact reason. Nevertheless, this does not significantly affect our results as $\text{O}_2$ absorbs primarily in the visible band where observations should be directed.

Thus, the presence of cloud cover impacts whether a robust detection of an atmospheric signal can be made within a reasonable exposure time. For example, a 24 hr long exposure is not enough to achieve net SNR = 5 for $\text{H}_2\text{O}$ on a cloudless planet but for a cloudy planet, the same exposure yields a net SNR $>$ 5. For both the cloudy and the cloud-free exo-Earth, comparing the net SNRs shows that $\text{O}_3$ is the easiest to detect followed by $\text{H}_2\text{O}$ and then $\text{O}_2$. However this is assuming that the instrument has a full bandpass and can simultaneously observe in all three wavelength bands. This might not be true for future missions with disparate wavelength coverage and narrow bandpasses. In this case, $\text{O}_2$ is the most promising candidate for detection as it has strong absorption line in the visible band. $\text{O}_3$ and $\text{H}_2\text{O}$ have broad absorption signals and may require multiple passes for a robust detection. We observe similar trends when considering other days as well, and hence this signal enhancement is not affected by the chosen day. 



\subsection{The impact of a global cloud layer at different pressure levels}

In this section, we explore the impact of a global cloud layer at different altitudes/pressure levels on the reflection spectra and the detectability of atmospheric gases. Similar work has been done before using 1d climate models (\cite{kawashima2019theoretical}) and we complement their results using a 3d empirical exo-Earth model. The thickness of the global cloud layer is kept constant in the pressure space (0.1 bar), and we consider 4 cases where this cloud layer is placed between different pressure levels in the atmosphere - (i) 0.8 - 0.7 bar, (ii) 0.48 - 0.38 bar, (iii) 0.34 - 0.24 bar and (iv) 0.18 - 0.08 bar. These are arbitrary pressure levels chosen according to the constraints of the 3D structure of the MERRA-2 data but we still try to sample the entire extent of the troposphere. The first three cases correspond to liquid water cloud layers while the last case corresponds to a water ice cloud layer, since temperatures between those pressure levels are low enough to form ice crystals. The cloud layer is `constructed' by converting all the gaseous water vapor in between the two pressure levels to cloud condensates. This conversion is done in two different ways - (i) we take a spatial average of the water vapor abundance (quantified by the water vapor mass fraction as given by the MERRA-2 data) over all latitudes and longitudes between the two pressure levels and convert this to a cloud mass fraction, or (ii) we divide the globe into $10^\circ$ latitude bins, and in each bin we take a spatial average of the water vapor mass fraction over all longitudes and latitudes and convert this to a cloud mass fraction, all within that bin. This is done to incorporate the spatial differences in water vapor over different parts of the globe. However, we find that the results of these two cases are nearly similar, since we cannot resolve spatial differences in cloud cover when the exo-Earth is 10 parsecs away. Thus, we only present results corresponding to the first case, with no latitudinal binning.  

\begin{figure*}
    \centering
    \includegraphics[width=0.95\textwidth]{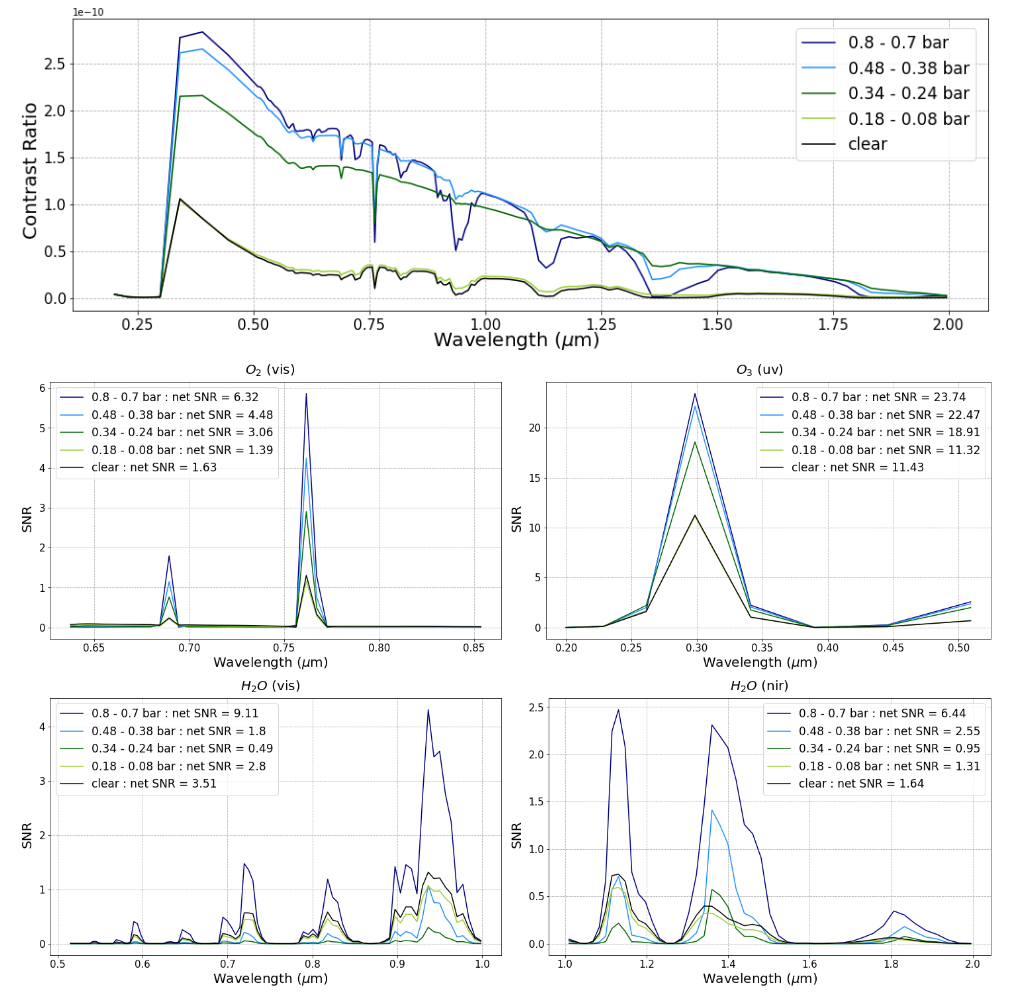}
    \caption{ 
    A global cloud layer of constant 0.1 bar thickness is considered at four different pressure levels - (i) 0.8 - 0.7 bar, (ii) 0.48 - 0.38 bar, (iii) 0.34 - 0.24 bar and (iv) 0.18 - 0.08 bar. The top panel shows the reflectance spectra for these four cases with a clear sky spectrum plotted for reference. As before, these simulations are done with a 6m HWO template with an exposure time of 24 hours for a planet that is 10 parsecs away. The middle panel shows the SNRs of $\text{O}_2$ in the visible band and $\text{O}_3$ in the UV band and the bottom panel shows the SNRs of $\text{H}_2\text{O}$ in the visible and NIR band. As the altitude of the cloud layer increases, the clouds become optically thinner and thus their reflectivity reduces leading to a lower continuum in the reflected light spectrum. The trends seen in the SNRs of all the three gases is a reflection of how they are distributed with respect to the global cloud layer.}
    \label{fig:cld_layers_at_diff_alts}
\end{figure*}

The top panel of figure \ref{fig:cld_layers_at_diff_alts} shows the reflectance spectra for an exo-Earth with a global cloud layer between 4 different pressure levels, as described above, with a clear sky spectrum also plotted for reference. The amount of water vapor decreases as we go to higher altitudes/lower pressures and subsequently the cloud layers become optically thinner. Thus, the reflected light continuum decreases as the altitude of the global cloud layer increases, but the continuum for all four cases is higher than the clear sky continuum. As mentioned before as well, the depth of the gaseous absorption signals is highly dependent on the distribution of clouds with respect to the vertical abundance of the gas. In addition to this, the water vapor absorption bands are also a function of how much water vapor has been converted to clouds between the considered pressure levels. 

The middle and bottom panels depict the SNRs of (i) $\text{O}_2$ in the visible band, (ii) $\text{O}_3$ in the UV band, and $\text{H}_2\text{O}$ in the (iii) visible and (iv) NIR band, with the net SNRs in the corresponding band given in the legend. For $\text{O}_2$, a zoomed-in plot depicting the two prominent absorption lines in the visible band is shown, however, the net SNRs are computed for the entire visible band. $\text{O}_2$ is a well-mixed gas and is present throughout the atmosphere with a constant mixing ratio of $\sim 0.21$. Low-lying and optically thick clouds significantly enhance the SNR for $\text{O}_2$ (compared to a clear sky) since a large fraction of $\text{O}_2$ is present above the cloud layer and contributes to absorption and moreover, the optically thick clouds increase the reflectivity of the disc. As the altitude of the cloud layer increases, it gets difficult to probe the atmosphere below the cloud layer and thus the amount of $\text{O}_2$ contributing to the absorption signal reduces which leads to a reduction in the SNR. $\text{O}_3$ is mainly abundant in the stratosphere, below the pressure levels being considered here, so the only impact of a cloud layer at higher pressures is to increase the reflectivity of the disc. As the altitude of the cloud layer increases, the clouds become optically thinner and less reflective, leading to a reduction in the SNRs for $\text{O}_3$. Water vapor is not a well-mixed gas and moreover, is also converted to cloud condensates between the pressure levels considered. These two factors contribute to the trend seen in the SNRs of $\text{H}_2\text{O}$ in the visible and the NIR band. We conclude that a low-lying optically thick cloud layer can significantly enhance the SNRs of well-mixed atmospheric constituents and thus reduce the exposure times required to make a robust detection. For $\text{O}_2$ and $\text{H}_2\text{O}$, the altitude of the global cloud layer dictates whether a 24 hr exposure yields net SNRs $\geq$ 5.

 \newpage

\subsection{The impact of naturally varying cloud cover}

In the previous sections, we quantified the effects of clouds by comparing atmospheres with the explicit presence and the absence of clouds as well as clouds at different altitudes in the atmosphere. In this section, we explore the impacts of the amount of cloud abundance on the reflectance spectra and the detectability of gases. The amount of cloud coverage is a highly variable quantity, impossible to predict and highly dependent on  climate and weather variability. The amount of cloud coverage can change significantly over different timescales and it is important to accurately quantify the impacts of these changes. Similar to Earth, we expect other terrestrial habitable worlds to have short-term weather patterns and long-term climate cycles resulting in a dynamic atmosphere which might impact the detectability of atmospheric constituents.


We go through the MERRA-2's historic cloud data  record to identify trends in cloud variability over different timescales.  
We specifically use MERRA-2 data which gives us simulated clouds, instead of other datasets with observed cloud data, like ISCCP (\cite{isccp_dataset}) or CERES (\cite{loeb2018clouds})) for the following reasons - (i) the MERRA-2 data is 3-dimensional and thus, the cloud data can be easily incorporated into our 3d exo-Earth model, (ii) MERRA-2 uses cloud mass fraction, in addition to cloud coverage as a proxy for a cloud abundance. This cloud mass fraction is essential for a radiative transfer calculation and it can be used to calculate an effective mass of cloud particles in the atmosphere. We find that a comparison between different datasets (for eg. MERRA-2 and ISCCP) is impractical because - (i) the MERRA-2 data is 3d while the ISCCP data is 2d and (ii) both datasets use different proxies for quantifying cloud abundances. The MERRA-2 pipeline simulates the cloud mass fraction and the areal cloud coverage at different layers in the atmosphere while the ISCCP algorithm converts observed reflected radiance to top-of-the atmosphere areal cloud coverage. A combination of simulated and observed cloud abundances might have yielded more accurate trends, however, we find that such a combination is not possible because of the different proxies employed. A more detailed comparison is needed to accurately integrate different datasets, however that is outside the scope of this work.    

We use the cloud mass fraction given by the MERRA-2 data to compute an effective global cloud mass. For an exo-Earth situated 10 parsecs away, we cannot resolve clouds at local scales and hence use this effective global cloud mass as a measure of global cloud abundance. Considering one column (lat $\times$ lon), we compute the pressure-weighted cloud mass fraction throughout that column or the liquid water path (LWP) as - 

\begin{equation}
\text{LWP} (\text{in} \hspace{0.1cm} \text{kg}/\text{m}^2) = \int_{\text{P}_{\text{s}}}^{0} \text{m}_{\text{c}} \, \frac{\text{dP}}{\text{g}} 
\end{equation}
where $\text{P}_{\text{s}}$ is the surface pressure, $\text{m}_{\text{c}}$ is the cloud mass fraction in each layer and g is the acceleration due to gravity. The area of a grid centered around latitude $\lambda$ and longitude $\phi$ is given by -

\begin{equation}
\text{A} = \text{r}_\oplus^2 \text{cos}(\lambda)\text{d}\lambda \text{d}\phi
\end{equation}
According to the resolution of the MERRA-2 data, $\text{d}\lambda = 0.5^\circ$ and $\text{d}\phi = 0.625^\circ$ everywhere on the grid. Multiplying the liquid water path with the area of the grid gives an effective cloud mass for one column. Taking a spatial average of this effective mass over all the columns gives the effective global cloud mass (in kgs). 

\begin{figure*}
    \centering
    \includegraphics[width=0.95\textwidth]{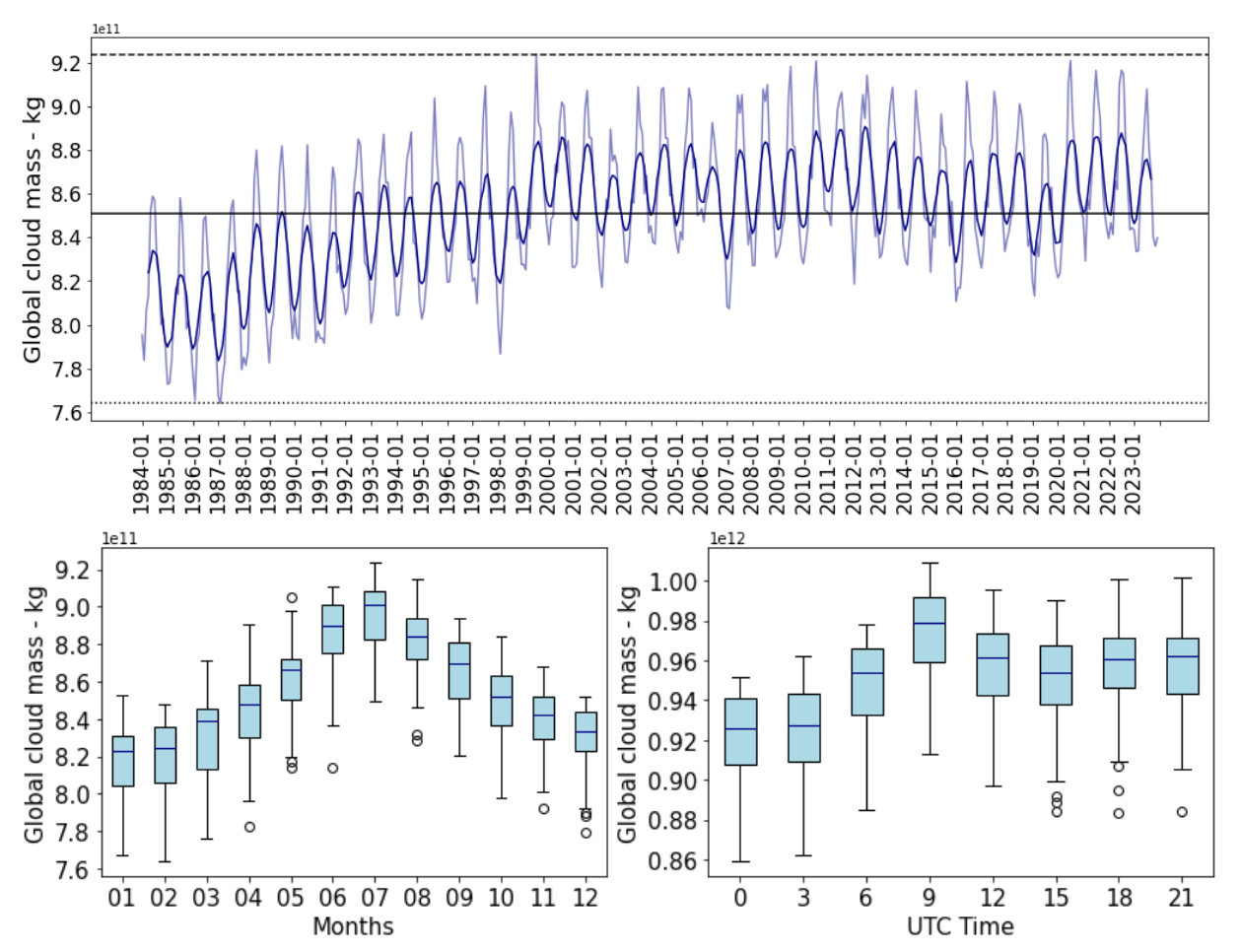}
    \caption{The top panel shows the evolution of the monthly averaged global cloud mass across MERRA-2's historic record from 1984-2023. The dashed and the dotted lines represent the maximum and minimum of the data which correspond to the months of July 1999 and February 1987 respectively. The solid line represents the mean which is closest to the January 2011 monthly averaged data. The faint blue curve represents the actual data while the dark blue curve is the rolling average over a window of 7 months. The bottom panels depict trends in cloud cover variability over seasonal and diurnal timescales. The left panel shows the trends in global cloud mass across an year and the boxes have been constructed using monthly averaged data from 1984-2023. The right panel depicts the diurnal trend and the boxes are plotted using instantaneous 3-hourly data for all days in the month of July 1999. For both these panels, the dark blue line represents the median, the edges of the box extend from the first quartile (Q1) to the third quartile (Q3), the whiskers extend to the farthest data points within 1.5 times the inter-quartile range (Q3 - Q1) from the box and the outliers are data points beyond this range.}
    \label{fig:cloud_cover_record}
\end{figure*}

\begin{figure*}
    \centering
    \includegraphics[width=0.95\textwidth]{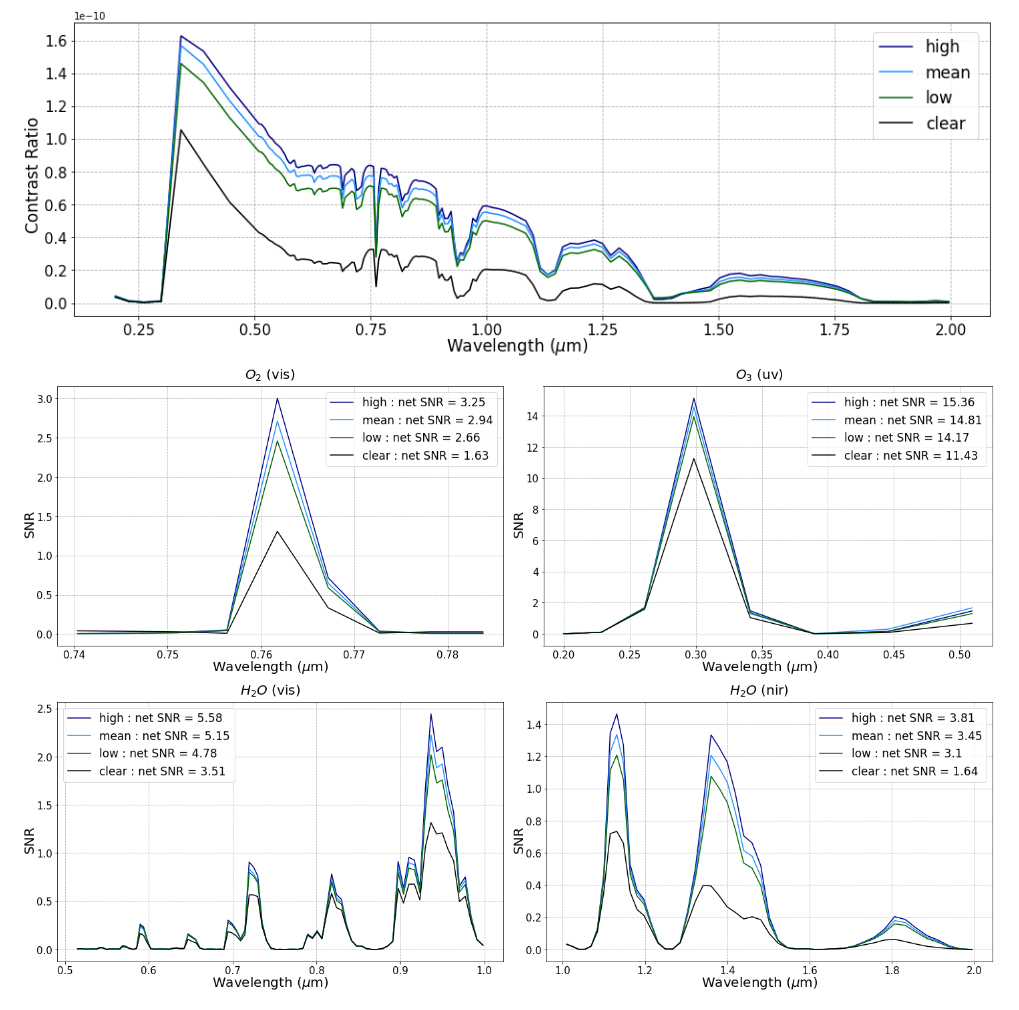}
    \caption{A comparison of the spectra and gaseous SNRs for an exo-Earth with varying amounts of cloud abundance. The high, mean, low cloud abundance labels correspond to exo-Earth atmospheres constructed using MERRA-2 data from the dates - 7th July 1999, 20th January 2011 and 15th February 1987 respectively. A clear sky scenario is also shown for reference. The simulations are done with a 6m HWO template with an exposure time of 24 hours considering that the planet is 10 pcs away and placed at quadrature. The reflected light continuum and the atmospheric SNRs increase with cloud abundance.}
    \label{fig:cloud_amt_comp}
\end{figure*}

The MERRA-2 data collection M2I3NVASM contains atmospheric data from 1984 - present, with a temporal resolution of 3 hours, and hence, it would be computationally intensive to go through all that data and identify trends in cloud cover.  MERRA-2 also has a separate collection M2TMNPCLD \cite{merra2_data_monthly_mean} that offers monthly averages of cloud data which we utilize to compute the monthly averaged effective global cloud mass from the years 1984 - 2023. Using the global cloud mass as a proxy, we identify the trends in global cloud cover at different time-scales. The top panel of figure \ref{fig:cloud_cover_record} shows the evolution of the global cloud mass across 1984-2023, where each data point represents a monthly average. The data shows evidence for an overall increase in the global cloud mass over the past few decades. The bottom left panel shows the seasonal cloud variability across an year. The boxes are drawn using the monthly averaged data from 1984-2023, where the median is depicted by the dark blue line. Cloud mass increases during the Northern summer months, peaking in July, and then reduces during the winter months. All the months show a $\sim 5-10\%$ variation in the monthly averaged cloud mass across the considered time period. For half the months, there are no outliers and the all data lies within the whiskers while the rest of the months show 1-3 outlier data points. This seasonal variability also underlies the long-term trend seen in the top panel. From this data, we identify the months corresponding to the highest, lowest and the mean global cloud mass (given by the dashed, solid and dotted lines respectively). July 1999 shows the highest monthly averaged global cloud mass - 9.24 $\times 10^{11}$ kg while February 1987 has the lowest cloud mass - 7.64 $\times 10^{11}$ kg. The mean cloud mass values of $8.5 \times 10^{11}$ kg is closest to the January 2011 data. Finally, the bottom right panel shows the diurnal cloud variability across the cloudiest month of July 1999. These boxes are plotted using data from the 3-hourly data collection M2I3NVASM. The diurnal variation is not significant and there are only a few outliers.

Having identified the months corresponding to the highest and the least monthly averaged global cloud mass, we use the 3-hourly data collection M2I3NVASM to compute the global cloud mass across all the days in those months and identify the days with the highest (7th July) and the lowest (15th February) cloud cover. Here, we have made the assumption that the month with the highest monthly averaged global cloud mass will have days with relatively high global cloud mass and vice versa and thus, we understand that these might not be the days with the absolute highest or lowest global cloud mass across MERRA-2's historic record. However, in the relative sense, these days can be classified as days with high and low cloud abundance respectively. Similarly, we also identify the day (20th January 2011) with the global cloud mass closest to the mean cloud mass across the historic record. We subsequently construct our exo-Earth atmospheres using empirical cloud data from these dates and analyze the impact of the amount of cloud cover on planetary spectra and SNRs of atmospheric gases. Note that instead of identifying individual days, we could have simply used the monthly averaged cloud data to construct our atmosphere but we didn't do so for two reasons - (i) the 3d structure of the monthly averaged data collection is different, it has 42 atmospheric layers, instead of the 72 layers in the 3-hourly data collection and (ii) analyzing data from individual days might reveal outliers with high weightage in the monthly average and this would yield a more dramatic comparison.  

The top panel of figure \ref{fig:cloud_amt_comp} shows the reflectance spectra for exo-Earth atmospheres constructed using MERRA-2 data from the following dates - 7th July 1999, 20th January 2011 and 15th February 1987. These cases are respectively labelled as high, mean and low, pertaining to the level of cloud abundance. A clear sky spectrum is also plotted for reference. The cloudy spectra show differences in the reflected light continuum which is attributed to the differences in the amount of cloud abundance and the subsequent disc albedo. These differences are prominent in the visible band where scattering and reflection by clouds is dominant. The spectra show a maximum variation of $\sim 15\%$ at $0.5\mu$m, which is much lower than the variations exhibited by a generic cloudy planet at different orbital phases. The middle and bottom panels depict the SNRs of atmospheric constituents $\text{O}_2$, $\text{O}_3$ and $\text{H}_2\text{O}$. $\text{O}_2$ is a well-mixed gas and thus an increase in the amount of cloud abundance and the subsequent albedo increases the SNRs. $\text{O}_3$ is dominant in the stratosphere and thus differences in tropospheric cloud coverage do not significantly alter its detectability. The SNRs of $\text{H}_2\text{O}$ also scale with cloud abundance with high cloud abundances leading to higher SNRs. Note that this scaling is due to an increase in the albedo for higher cloud abundances and not because of an increased absorption by water cloud condensates. These condensates absorb at much higher wavelengths in the infrared (resulting in the greenhouse effect), which is not relevant here as this is beyond our considered wavelength bands. We conclude that atmospheres with higher cloud abundance increase the detectability of the atmospheric gases considered here and also reduce the exposure times required to make robust detections of these gases. This is especially highlighted by comparing the net SNRs of $\text{H}_2\text{O}$ in the visible band. Setting a threshold of SNR=5, its evident that 24 hours is not enough to make a robust detection of $\text{H}_2\text{O}$ in the visible band for an atmosphere with low cloud abundance. Meanwhile, atmospheres with mean and high cloud abundances result in net SNRs $>$ 5.

We note that even though the data used to construct the cloud distributions considered here span several decades, the differences presented in figure \ref{fig:cloud_amt_comp} are not meant to representative of variations on observationally relevant timescales. We simply take advantage of MERRA-2's rich historical record to find periods with significant differences in cloud amount and use the subsequent empirical data to construct realistic cloud distributions on an exo-Earth. It's impossible to have any prior information on the amount of cloud abundance and hence during an observation, an exo-Earth might be highly cloudy, less cloudy or somewhere in between the two extremes. Figure \ref{fig:cloud_amt_comp} shows the possible variations in spectra and atmospheric detectability for an exo-Earth with variable cloud abundance and highlights the fact that it is difficult to establish a stable atmospheric baseline in the presence of clouds.

For cases when we don't have access to the full reflection continuum (for example, if observations are restricted to certain bandpasses), our analysis also highlights a possible degeneracy between the amount of cloud abundance and the abundance of an atmospheric gas while interpreting observed SNRs. For eg. an atmosphere with a high cloud abundance and low $\text{O}_2$ concentration might yield the same SNRs as an atmosphere with high $\text{O}_2$ concentration but low cloud abundance. Exploring this degeneracy in detail and possible solutions to resolve it requires an in-depth analysis using photochemical modelling and retrievals, which is outside the scope of this study.


\subsection{Comparison between different mission concepts}

\begin{figure*}
    \centering
    \includegraphics[width=0.95\textwidth]{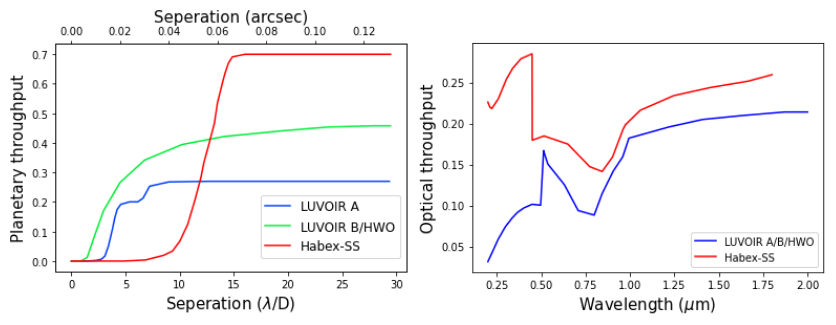}
    \caption{The planetary and optical throughputs for all the four instruments considered for the comparison.}
    \label{fig:thrput}
\end{figure*}

In this section, We do a comparative analysis of future direct-imaging mission concepts - (i) LUVOIR-A (15m), (ii) LUVOIR-B (8m), (iii) HWO (6m) and (iv) Habex with starshade (4m), in investigating the impacts of clouds. We use PSG to simulate all our observations with these instruments and all the instrument parameters, as shown in table \ref{table:mis_con_par}, are taken from the respective final reports (\cite{luvoir2019luvoir,gaudi2020habitable}). Since the Habitable Worlds Observatory report is still under development, we consider the HWO instrument parameters to be the same as the LUVOIR-B parameters, except with a smaller mirror size. The coronagraphic planetary throughput and the optical throughput of each instrument is shown in figure \ref{fig:thrput}. For a detailed discussion on the throughputs and the PSG noise model, refer to \cite{checlair2021probing} or the PSG handbook \cite{psg_handbook}. 

\movetableright=-0.7in
\begin{table*}
    \centering
    \begin{tabular}{m{2.5cm}m{3cm}m{3cm}m{3cm}m{3cm}}
    \hline
    \textbf{Parameter} & \textbf{LUVOIR-A} & \textbf{LUVOIR-B} & \textbf{Habex/SS} & \textbf{HWO} \\
    \hline
    Diameter & 15m & 8m & 4m & 6m\\
    \hline
    Spectral range & UV: 0.2 - 0.515 $\mu$m & UV: 0.2 - 0.515 $\mu$m & UV: 0.2 - 0.45$ \mu$m & UV: 0.2 - 0.515 $\mu$m\\
     & VIS: 0.515 - 1 $\mu$m & VIS: 0.515 - 1 $\mu$m & VIS: 0.45 - 0.975 $\mu$m & VIS: 0.515 - 1 $\mu$m\\
     & NIR: 1 - 2 $\mu$m & NIR: 1 - 2 $\mu$m & NIR: 0.975 - 1.8 $\mu$m & NIR: 1 - 2 $\mu$m  \\
    \hline
    Resolution & UV: 7 RP & UV: 7 RP & UV: 7 RP & UV: 7 RP\\ & VIS: 140 RP & VIS: 140 RP & VIS: 140 RP & VIS: 140 RP\\ & NIR: 70 RP & NIR: 70 RP & NIR: 40 RP & NIR: 70 RP\\
    \hline
    Exozodi level & 4.5 & 4.5 & 4.5 & 4.5\\ 
    \hline
    Contrast & 1 $\times 10^{-10}$ & 1 $\times 10^{-10}$ & 1 $\times 10^{-10}$ & 1 $\times 10^{-10}$\\
    \hline
    IWA & 4 $\lambda$/D & 3.5 $\lambda$/D & UV: 39 mas & 3.5 $\lambda$/D \\ & & & VIS: 58 mas & \\ & & & NIR: 104 mas &\\
    \hline
    Read noise & UV: 0 & UV: 0 & UV: 0.008 & UV: 0\\ & VIS: 0 & VIS: 0 & VIS: 0.008 & VIS: 0 \\ & NIR: 2.5 & NIR: 2.5 & NIR: 0.32 & NIR: 2.5\\ 
    \hline
    Dark noise & UV: 3 $\times 10^{-5}$ & UV: 3 $\times 10^{-5}$ & UV: 3 $\times 10^{-5}$ & UV: 3 $\times 10^{-5}$\\ & VIS: 3 $\times 10^{-5}$ & VIS: 3 $\times 10^{-5}$ & VIS: 3 $\times 10^{-5}$ & VIS: 3 $\times 10^{-5}$ \\ & NIR: 0.002 & NIR: 0.002 & NIR: 0.005 & NIR: 0.002\\
    \hline
    \end{tabular}
    \caption{The parameters that describe the instrumental configuration for the four mission concepts, LUVOIR A/B, Habex with SS, and HWO. Parameter values adapted from the respective final reports. We assume that the HWO as the same instrumental parameters as LUVOIR-B, except for a smaller mirror size.}
    \label{table:mis_con_par}    

\end{table*}

We simulate observations with all the instruments considering that our exo-Earth is placed 10 parsecs away, situated at quadrature and set our exposure times to 24 hrs. As described above, we take an average of the atmospheric data over all parts of the disc to represent the observation of an integrated disc. We compute the individual wavelength-dependent SNRs and the band-wise net SNRs of the three gases - $\text{O}_2$, $\text{H}_2\text{O}$ and $\text{O}_3$. Following the method given in \cite{checlair2021probing}, we then calculate the exposure times required to achieve SNRs = 5.

\begin{equation}
    \text{t}_{\text{exp}} = 24 (\text{hr}) \times \Big (\frac{5}{\text{SNR}} \Big)^2
\end{equation}

For this comparison, we consider six types of atmospheres - an atmosphere having patchy clouds with (i) high, (ii) mean and (iii) low cloud abundance, an atmosphere with global cloud coverage with clouds between two pressure levels, (iv) 0.8 - 0.7 bar and (v) 0.34 - 0.24 bar and a (vi) clear atmosphere with no clouds. The atmospheres with varying cloud abundances are constructed using empirical MERRA-2 data corresponding to those dates, as described in the previous section. To compute these SNRs and exposure times, we consider the strongest absorption lines/bands - (i) the $\text{O}_2$ A line at $\sim 0.76 \mu$m in the visible band, (ii) the $\text{H}_2\text{O}$ absorption band across 0.9-1 $\mu$m in the visible band, (iii) the $\text{H}_2\text{O}$ absorption band across 1.25-1.6 $\mu$m in the NIR band and, (iv) the $\text{O}_3$ absorption band across 0.2-0.4 $\mu$m in the UV band. 

For the absorption bands, we compute the net SNR across the range of absorption and then ingest this net SNR into the equation above to calculate the exposure time. Tables \ref{table:o2_exp_time}, \ref{table:h2o_vis_exp_time}, \ref{table:h2o_nir_exp_time} and \ref{table:o3_exp_time} depict the exposure times required to make a robust detection defined by setting a threshold SNR=5 for all the six atmospheres described above considering four future direct-imaging mission concepts. As shown in \autoref{table:mis_con_par} and fig. \ref{fig:thrput}, the main differences between these mission concepts that affect the exposure time computations are the instrument mirror size and the instrument throughputs. The instrumental noise increases with reducing mirror size causing a subsequent increase in the required exposure times. Thus, LUVOIR-A with the largest mirror size of 15m shows the highest performance and yields the lowest exposure times for all absorbing gases. Its followed by LUVOIR-B with an 8m big primary mirror. However, Habex-SS (4m) performs better than HWO (6m) due to its superior planetary and optical throughput (see \autoref{fig:thrput}) in spite of a smaller mirror size.

\movetableright=-0.7in
\begin{table*}[h!]
    \centering
    \begin{tabular}{ccccc}
    \hline
    \textbf{Type of Atmosphere} & \textbf{LUVOIR-A} & \textbf{LUVOIR-B} & \textbf{HWO} & \textbf{Habex-SS} \\
    \hline
    Patchy clouds with high cloud abundance & 2.4 hr & 18.24 hr & 66.54 hr & 63.79 hr \\
    \hline
    Patchy clouds with mean cloud abundance & 2.91 hr & 22.33 hr & 81.62 hr & 78.32 hr\\
    \hline
    Patchy clouds with low cloud abundance & 3.49 hr & 27.05 hr & 99.13 hr & 95.32 hr \\
    \hline
    Global cloud layer from 0.8 - 0.7 bar & 0.73 hr & 4.97 hr & 17.47 hr & 16.31 hr\\
    \hline
    Global cloud layer from 0.34 - 0.24 bar & 3.23 hr & 20.7 hr & 71.17 hr & 65.68 hr\\
    \hline
    No clouds & 10.94 hr & 93.14 hr & 350.65 hr & 343.02 hr\\
    \hline
    \end{tabular}
    \caption{The exposure times required to make a robust detection (SNR = 5) of the $\text{O}_2$ A line for the six types of atmospheres described above for an exo-Earth at 10 parsecs away placed at quadrature in its orbit.}
    \label{table:o2_exp_time}    

\end{table*}

\movetableright=-0.7in
\begin{table*}[h!]
    \centering
    \begin{tabular}{ccccc}
    \hline
    \textbf{Type of Atmosphere} & \textbf{LUVOIR-A} & \textbf{LUVOIR-B} & \textbf{HWO} & \textbf{Habex-SS} \\
    \hline
    Patchy clouds with high cloud abundance & 0.59 hr & 5.43 hr & 22.34 hr & 16.77 hr \\
    \hline
    Patchy clouds with mean cloud abundance & 0.69 hr & 6.36 hr & 26.21 hr & 19.68 hr\\
    \hline
    Patchy clouds with low cloud abundance & 0.8 hr & 7.47 hr & 30.95 hr & 23.16 hr \\
    \hline
    Global cloud layer from 0.8 - 0.7 bar & 0.26 hr & 2.08 hr & 8.23 hr & 6.08 hr\\
    \hline
    Global cloud layer from 0.34 - 0.24 bar & 94.6 hr & 692.12 hr & 2645.57 hr & 1992.9 hr\\
    \hline
    No clouds & 1.3 hr & 13.81 hr & 59.05 hr & 44.35 hr\\
    \hline
    \end{tabular}
    \caption{The exposure times required to make a robust detection (SNR = 5) of the $\text{H}_2\text{O}$ 0.9-1 $\mu$m absorption band for the six types of atmospheres for an exo-Earth placed 10 parsecs away at quadrature.}
    \label{table:h2o_vis_exp_time}    

\end{table*}

\movetableright=-0.7in
\begin{table*}[h!]
    \centering
    \begin{tabular}{ccccc}
    \hline
    \textbf{Type of Atmosphere} & \textbf{LUVOIR-A} & \textbf{LUVOIR-B} & \textbf{HWO} & \textbf{Habex-SS} \\
    \hline
    Patchy clouds with high cloud abundance & 0.44 hr & 7.75 hr & 74.83 hr & 18.85 hr \\
    \hline
    Patchy clouds with mean cloud abundance & 0.55 hr & 9.67 hr & 93.23 hr & 24.16 hr\\
    \hline
    Patchy clouds with low cloud abundance & 0.71 hr & 12.29 hr & 118.36 hr & 31.34 hr \\
    \hline
    Global cloud layer from 0.8 - 0.7 bar & 0.15 hr & 2.58 hr & 24.94 hr & 6.1 hr\\
    \hline
    Global cloud layer from 0.34 - 0.24 bar & 5.29 hr & 81 hr & 757.95 hr & 647.04 hr\\
    \hline
    No clouds & 4.56 hr & 79.98 hr & 756.44 hr & 230.95 hr\\
    \hline
    \end{tabular}
    \caption{The exposure times required to make a robust detection (SNR = 5) of the $\text{H}_2\text{O}$ 1.25-1.6 $\mu$m absorption band for the six types of atmospheres described above. Exo-Earth is placed 10 parsecs away at quadrature.}
    \label{table:h2o_nir_exp_time}    

\end{table*}

\movetableright=-0.7in
\begin{table*}[h!]
    \centering
    \begin{tabular}{ccccc}
    \hline
    \textbf{Type of Atmosphere} & \textbf{LUVOIR-A} & \textbf{LUVOIR-B} & \textbf{HWO} & \textbf{Habex-SS} \\
    \hline
    Patchy clouds with high cloud abundance & 0.23 hr & 0.9 hr & 2.57 hr & 1.22 hr \\
    \hline
    Patchy clouds with mean cloud abundance & 0.25 hr & 0.97 hr & 2.78 hr & 1.31 hr\\
    \hline
    Patchy clouds with low cloud abundance & 0.27 hr & 1.05 hr & 3.01 hr & 1.43 hr \\
    \hline
    Global cloud layer from 0.8 - 0.7 bar & 0.1 hr & 0.38 hr & 1.08 hr & 0.51 hr\\
    \hline
    Global cloud layer from 0.34 - 0.24 bar & 0.15 hr & 0.59 hr & 
    1.7 hr & 0.8 hr\\
    \hline
    No clouds & 0.41 hr & 1.61 hr & 4.61 hr & 2.17 hr\\
    \hline
    \end{tabular}
    \caption{The exposure times required to make a robust detection (SNR = 5) of the $\text{O}_3$ 0.2-0.4 $\mu$m band for the six types of atmospheres.}
    \label{table:o3_exp_time}    

\end{table*}


For a given instrument, the trend in the exposure times highly depends on the presence, amount and distribution of cloud cover in the atmosphere, as well as on the distribution of the absorbing gas. The presence of a global cloud layer at low altitudes ensures that a dominant fraction of any absorbing gas is present above the cloud layer and it simply enhances the reflectivity of the disc. Thus, an atmosphere with a global cloud layer between 0.8 - 0.7 bar leads to the highest SNRs and the lowest exposure times required to make a robust detection. In contrast, the presence of a global cloud layer at high altitudes makes it difficult to probe the atmosphere below resulting in low SNRs and higher exposure times. This difference is especially highlighted for observations of the $\text{H}_2\text{O}$ 0.9-1 $\mu$m signal (see \autoref{table:h2o_vis_exp_time}), where, depending on the instrument, it takes $\sim 330-360$ times longer to make an SNR=5 detection for an atmosphere with a high-altitude global cloud layer. Depending on the absorbing gas, either an atmosphere with a global high altitude cloud layer or a clear atmosphere yield the lowest SNRs and the highest exposure times. Note that our $\text{O}_2$ exposure times for the clear sky scenario are higher than those computed in \cite{checlair2021probing}, as they assume simultaneous wavelength coverage and calculate the net SNR in all the three wavelength bands. In contrast, we only focus on the $\text{O}_2$ A line in the visible band. Similarly, the difference in the $\text{O}_3$ exposure times might be additionally caused by changes in the sampling algorithm of PSG (\cite{Saxena_2021}).

Since we expect an abundance of patchy clouds on exo-Earth atmospheres, the cloud-free case is a hypothetical scenario and not a baseline assumption. The three cases with varying levels of patchy cloud abundance are more realistic, and hence a detailed comparison amongst them is necessary. The $\text{O}_3$ SNRs are not significantly affected by the amount of tropospheric patchy clouds, and hence show little variation. In contrast the SNRs of $\text{O}_2$ and $\text{H}_2\text{O}$ show a significant correlation with the amount of patchy clouds, as shown by the computed exposure times required to yield SNR = 5. Thus, depending on the atmospheric constituent, the amount of patchy clouds result in a $15-60 \%$ variation in the exposure times for an HWO 6m instrument.  These variations would be confounded if the instrument doesn't have simultaneous wavelength coverage. For instance, HWO will take 143 hrs to detect all three gases with SNR = 5 if the atmosphere has high cloud abundance versus 220 hrs if the atmosphere has low cloud abundance, a $54 \%$ increase in exposure time. This has significant implications for observing programs dedicated to accurately characterizing exo-Earth atmospheres.

\section{Discussion and conclusions}

To summarize, we construct an accurate 3D model of an exo-Earth using empirical 3D atmospheric data from the MERRA-2 reanalysis package. We quantify the effects of clouds and cloud variability on the reflectance spectra and SNRs of key molecules by simulating observations of this exo-Earth with the GlobES module of PSG for atmospheres with different cloud distributions and abundances. Clouds enhance the continuum in the reflected light by increasing the albedo of the disc. Different amounts of cloud cover over the visible portions of the planet's disc cause variations in the spectra of a rotating planet. These variations are more significant than the ones caused by the changing surface albedo of the visible disc. \autoref{fig:rot_exoearth_spec} shows that the spectra of a cloudy planet show at max a $\sim 75\%$ variation in reflected flux at 0.5 $\mu$m for different rotational phases. In contrast, the cloud-free spectra only vary by $\sim 15 \%$ at the same wavelength, due to the changing surface albedo. Depending on total cloud coverage and their vertical position, patchy clouds can increase or decrease the SNRs of all three atmospheric gases considered here, $\text{O}_2$, $\text{H}_2\text{O}$ and $\text{O}_3$ and thus enhance/diminish their detectability in exo-Earth atmospheres. The impact of cloud variability can potentially confound efforts to retrieve a stable baseline atmosphere for a planet - this is inherently because such a baseline is a physical approximation.  Similar to transit studies of exoplanets, cloud variability and weather may have chromatic effects on planetary spectra (\cite{Powell_2019}) that may limit (\cite{Fauchez_2019}) or enhance detection of key features depending on their magnitude relative to noise (\cite{May_2021}) and other factors. Importantly, observations of the Earth utilized in an `Earth as an Exoplanet' manner can play a key role in informing how such variability may play these roles depending on different relevant temporal scales.

The impact of patchy clouds and weather variability for direct imaging is relevant to the most proximate missions aimed at detecting and characterizing potential exo-Earths. The Habitable Worlds Observatory's main goal is to identify and directly image at least 25 potentially habitable worlds - this target is based upon statistical yield calculations that require assumptions about both the astrophysical scene and planetary properties. Some of these planets are expected to have large fractions of water on the surface, and like the Earth, consequent abundant cloud cover.Cloud cover variability is likely to impact \textit{both} detection and characterization of these worlds, thus influencing attainment of the required yield over a nominal mission design.  Since cloud cover difference and the variability that may drive it are most apparent in between 0.3 to 0.7 $\mu$m in the visible, this will likely have pronounced impacts on detection as recent work has found an optimum detection wavelength of 0.5 $\mu$m for the majority of targets relevant to HWO (\cite{10.1117/1.JATIS.10.1.014005}).  Similarly, the non-monotonic relationships in SNR changes due to cloud variability and height for oxygen and water, suggest that optimal wavelengths for detection of some key biomarkers are also likely to be impacted in complicated ways by such variability. Additional work exploring the general impact such variability may have on detection and characterization of a potential HWO sample of habitable planet candidates is important precursor work for an observation strategy.  Critically, the mission plan may also require an understanding of the impact of weather variability on the observational budget required to meet key objectives.  Given the 20-50\% variation in times to an SNR=5 signal for key molecules in tables \ref{table:o2_exp_time} - \ref{table:o3_exp_time}, the budget required to achieve detection and characterization of a sufficient sample to achieve mission objectives should account for this uncertainty.

We note that precisely quantifying the effect of clouds on SNRs requires an understanding of the vertical distribution of clouds with respect to the vertical distribution of the atmospheric constituents. For example, $\text{O}_2$ and $\text{H}_2\text{O}$ are abundant in the troposphere while $\text{O}_3$ is dominant much higher in the stratosphere. Thus, the differences in their SNR enhancements in the presence of clouds are a reflection of this distribution. We show that atmospheres with low-lying thick clouds produce the highest gaseous SNRs as such clouds greatly increase the albedo of the disc and the low altitude of the cloud layer ensures that a majority of the atmospheric scale height is probed. On the other hand, high-altitude clouds make it difficult to probe the atmosphere below them and thus reduce the detectability of any gases abundant there.  While observational constraints are unlikely to be obtained by HWO or other proximate planned direct imaging missions, variability due to varying cloud coverage may be a novel way of constraining cloud altitude in some limited cases. 

Using MERRA-2's historic record, we have shown that Earth shows significant variation in cloud abundance across different time-scales. On average, Earth's total cloud abundance has increased over the past few decades. On an annual scale, Earth's cloud cover peaks in the norther summer months of June - August and subsequently decreases during the winter months. We expect any exo-Earth with some obliquity to show a seasonal pattern in cloud variability. This implies that the planet will have higher cloud abundance for a certain duration of its orbital period. Thus, there is an optimum period to observe such a planet if the goal is to detect atmospheric constituents and obtain sufficiently high gaseous SNRs within a reasonable exposure time. However, to produce the highest observational yields, this optimum period should fortuitously coincide with the period when the planet it actually observable (within the bounds of the inner and outer working angle of the coronagraph and at an optimum orbital phase). Thus, observation can underestimate the SNRs, especially if the planet has low seasonal cloud cover during observation. This is further complicated by a possible degeneracy between the amount of cloud abundance and the concentration of a gases while interpreting observed SNRs. In addition to seasonal cloud variation, terrestrial planets may also exhibit diurnal cloud variations due to rotation (given that they are sufficiently far away from their host star to escape tidal effects). If the planet is under observation for much longer than its rotational period, these diurnal variations will be averaged out. Taking multiple small exposures over an extended observation period might reveal periodic trends in planetary spectra and SNRs caused by the diurnal variation in cloud cover, and such trends can also be used to estimate the rotation period of the planet. There are however a number of caveats in leveraging cloud variability to optimize observational strategy. First, we have only examined an initial general impact of water cloud coverage on the atmosphere - other factors such as variable hazes (such as those generated by volcanic eruptions) may also influence observations in transient ways.  Variability signatures may or may not exceed noise levels for different systems, and the impact in both cases on retrieved parameters should be examined. Finally, variability due to sources external to the planet that are difficult to extricate from weather variability signatures such as time-varying signatures of exozodiacal dust (\cite{2012_Roberge_Dust, 2012_Defrere_Dust} or the presence of additional bodies in the system (\cite{Saxena_2022}) may also be degenerate and confounding factors. However, if the planets we target to directly image in the future are hoped to be Earth-like, we should expect cloud variability based upon our observations of the present-day Earth, and should consider and incorporate both the challenges and potential opportunity such variability may provide.  



\begin{acknowledgements}

We would like to thank Vincent Kofman, Geronimo Villanueva and Allison Payne for conversations that helped improve the quality of this paper. The authors also acknowledge support from the Goddard Space Flight Center (GSFC) Sellers Exoplanet Environments Collaboration (SEEC), which is supported by the NASA Planetary Science Division's Research Program. This material is based upon work supported by NASA under award number 80GSFC21M0002. The authors acknowledge the support and the resources provided by PARAM Brahma Facility under the National Supercomputing Mission, Government of India at the Indian Institute of Science Education and Research, Pune.   

\end{acknowledgements}

\section{Appendix A: Configuration files}

The configuration files can be uploaded to the web interface or used with the API for PSG to simulate spectra, and are available in the manuscript metadata. There are four configuration files corresponding to the MERRA-2 data for four dates - 1st July 2000 (randomly selected date for calibration), 15th February 1987 (low patchy cloud abundance), 20th Janunary 2011 (closest to mean patchy cloud abundance) and 7th July 1999 (high patchy cloud abundance). Two files correspond to atmospheres with a low-altitude and a high-altitude global cloud layer.

\newpage
\bibliography{refs}{}

\begin{thebibliography}{}
\expandafter\ifx\csname natexlab\endcsname\relax\def\natexlab#1{#1}\fi
\providecommand{\url}[1]{\href{#1}{#1}}
\providecommand{\dodoi}[1]{doi:~\href{http://doi.org/#1}{\nolinkurl{#1}}}
\providecommand{\doeprint}[1]{\href{http://ascl.net/#1}{\nolinkurl{http://ascl.net/#1}}}
\providecommand{\doarXiv}[1]{\href{https://arxiv.org/abs/#1}{\nolinkurl{https://arxiv.org/abs/#1}}}

\bibitem[{Arnold {et~al.}(2002)Arnold, Gillet, Lardière, Riaud, \& Schneider}]{Arnold_2002}
Arnold, L., Gillet, S., Lardière, O., Riaud, P., \& Schneider, J. 2002, Astronomy \& Astrophysics, 392, 231–237

\bibitem[{Brooke {et~al.}(1998)Brooke, Knacke, Encrenaz, Drossart, Crisp, \& Feuchtgruber}]{brooke1998models}
Brooke, T., Knacke, R., Encrenaz, T., {et~al.} 1998, Icarus, 136, 1

\bibitem[{Buras {et~al.}(2011)Buras, Dowling, \& Emde}]{buras2011new}
Buras, R., Dowling, T., \& Emde, C. 2011, Journal of Quantitative Spectroscopy and Radiative Transfer, 112, 2028

\bibitem[{Cahoy {et~al.}(2010)Cahoy, Marley, \& Fortney}]{Cahoy_2010}
Cahoy, K.~L., Marley, M.~S., \& Fortney, J.~J. 2010, The Astrophysical Journal, 724, 189–214

\bibitem[{Checlair {et~al.}(2021)Checlair, Villanueva, Hayworth, Olson, Komacek, Robinson, Popovi{\'c}, Yang, \& Abbot}]{checlair2021probing}
Checlair, J.~H., Villanueva, G.~L., Hayworth, B.~P., {et~al.} 2021, The Astronomical Journal, 161, 150

\bibitem[{{Defr{\`e}re} {et~al.}(2012){Defr{\`e}re}, {Stark}, {Cahoy}, \& {Beerer}}]{2012_Defrere_Dust}
{Defr{\`e}re}, D., {Stark}, C., {Cahoy}, K., \& {Beerer}, I. 2012, in Society of Photo-Optical Instrumentation Engineers (SPIE) Conference Series, Vol. 8442, Space Telescopes and Instrumentation 2012: Optical, Infrared, and Millimeter Wave, ed. M.~C. {Clampin}, G.~G. {Fazio}, H.~A. {MacEwen}, \& J.~{Oschmann}, Jacobus~M., 84420M

\bibitem[{Edwards(1992)}]{edwards_genln2}
Edwards, D. 1992, NCAR/TN-367-STR, National Center for Atmospheric Research, Boulder, Co.

\bibitem[{Fauchez {et~al.}(2019)Fauchez, Turbet, Villanueva, Wolf, Arney, Kopparapu, Lincowski, Mandell, de~Wit, Pidhorodetska, Domagal-Goldman, \& Stevenson}]{Fauchez_2019}
Fauchez, T.~J., Turbet, M., Villanueva, G.~L., {et~al.} 2019, The Astrophysical Journal, 887, 194

\bibitem[{Friedl \& Sulla-Menashe(2015)}]{modis_data}
Friedl, M., \& Sulla-Menashe, D. 2015, NASA EOSDIS Land Processes Distributed Active Archive Center

\bibitem[{Gao {et~al.}(2021)Gao, Wakeford, Moran, \& Parmentier}]{gao2021aerosols}
Gao, P., Wakeford, H.~R., Moran, S.~E., \& Parmentier, V. 2021, Aerosols in exoplanet atmospheres,  Wiley Online Library

\bibitem[{Gaudi {et~al.}(2021)Gaudi, Meyer, \& Christiansen}]{Gaudi_2021}
Gaudi, B.~S., Meyer, M., \& Christiansen, J. 2021, The Demographics of Exoplanets (IOP Publishing)

\bibitem[{Gaudi {et~al.}(2020)Gaudi, Seager, Mennesson, Kiessling, Warfield, Cahoy, Clarke, Domagal-Goldman, Feinberg, Guyon, Kasdin, Mawet, Plavchan, Robinson, Rogers, Scowen, Somerville, Stapelfeldt, Stark, Stern, Turnbull, Amini, Kuan, Martin, Morgan, Redding, Stahl, Webb, Alvarez-Salazar, Arnold, Arya, Balasubramanian, Baysinger, Bell, Below, Benson, Blais, Booth, Bourgeois, Bradford, Brewer, Brooks, Cady, Caldwell, Calvet, Carr, Chan, Cormarkovic, Coste, Cox, Danner, Davis, Dewell, Dorsett, Dunn, East, Effinger, Eng, Freebury, Garcia, Gaskin, Greene, Hennessy, Hilgemann, Hood, Holota, Howe, Huang, Hull, Hunt, Hurd, Johnson, Kissil, Knight, Kolenz, Kraus, Krist, Li, Lisman, Mandic, Mann, Marchen, Marrese-Reading, McCready, McGown, Missun, Miyaguchi, Moore, Nemati, Nikzad, Nissen, Novicki, Perrine, Pineda, Polanco, Putnam, Qureshi, Richards, Riggs, Rodgers, Rud, Saini, Scalisi, Scharf, Schulz, Serabyn, Sigrist, Sikkia, Singleton, Shaklan, Smith, Southerd, Stahl, Steeves, Sturges, Sullivan, Tang, Taras,
  Tesch, Therrell, Tseng, Valente, Buren, Villalvazo, Warwick, Webb, Westerhoff, Wofford, Wu, Woo, Wood, Ziemer, Arney, Anderson, Maíz-Apellániz, Bartlett, Belikov, Bendek, Cenko, Douglas, Dulz, Evans, Faramaz, Feng, Ferguson, Follette, Ford, García, Geha, Gelino, Götberg, Hildebrandt, Hu, Jahnke, Kennedy, Kreidberg, Isella, Lopez, Marchis, Macri, Marley, Matzko, Mazoyer, McCandliss, Meshkat, Mordasini, Morris, Nielsen, Newman, Petigura, Postman, Reines, Roberge, Roederer, Ruane, Schwieterman, Sirbu, Spalding, Teplitz, Tumlinson, Turner, Werk, Wofford, Wyatt, Young, \& Zellem}]{gaudi2020habitable}
Gaudi, B.~S., Seager, S., Mennesson, B., {et~al.} 2020, The Habitable Exoplanet Observatory (HabEx) Mission Concept Study Final Report.
\newblock \doarXiv{2001.06683}

\bibitem[{Gelaro {et~al.}(2017)Gelaro, McCarty, Su{\'a}rez, Todling, Molod, Takacs, Randles, Darmenov, Bosilovich, Reichle, {et~al.}}]{gelaro2017modern}
Gelaro, R., McCarty, W., Su{\'a}rez, M.~J., {et~al.} 2017, Journal of climate, 30, 5419

\bibitem[{{Global Modeling and Assimilation Office (GMAO)}(2015{\natexlab{a}})}]{merra2_data_citation}
{Global Modeling and Assimilation Office (GMAO)}. 2015{\natexlab{a}}, {MERRA-2 inst3\_3d\_asm\_Nv: 3d,3-Hourly,Instantaneous, Model-Level, Assimilation, Assimilated Meteorological Fields V5.12.4,Greenbelt, MD, USA, Goddard Earth Sciences Data and Information Services Center (GES DISC), Accessed: 2023-11-24, 10.5067/SUOQESM06LPK}

\bibitem[{{Global Modeling and Assimilation Office (GMAO)}(2015{\natexlab{b}})}]{merra2_data_monthly_mean}
---. 2015{\natexlab{b}}, {MERRA-2 tavgM\_3d\_cld\_Np: 3d,Monthly mean,Time-Averaged,Pressure-Level,Assimilation,Cloud Diagnostics V5.12.4, Greenbelt, MD, USA, Goddard Earth Sciences Data and Information Services Center (GES DISC), Accessed: 2023-11-24, 10.5067/J9R0LXGH48JR}

\bibitem[{Gordon {et~al.}(2022)Gordon, Rothman, Hargreaves, Hashemi, Karlovets, Skinner, Conway, Hill, Kochanov, Tan, {et~al.}}]{gordon2022hitran2020}
Gordon, I.~E., Rothman, L.~S., Hargreaves, R., {et~al.} 2022, Journal of quantitative spectroscopy and radiative transfer, 277, 107949

\bibitem[{Grenfell {et~al.}(2007)Grenfell, Stracke, von Paris, Patzer, Titz, Segura, \& Rauer}]{grenfell2007response}
Grenfell, J.~L., Stracke, B., von Paris, P., {et~al.} 2007, Planetary and Space Science, 55, 661

\bibitem[{Gu {et~al.}(2021)Gu, Fan, Li, Bartlett, Natraj, Jiang, Crisp, Hu, Tinetti, \& Yung}]{gu2021earth}
Gu, L., Fan, S., Li, J., {et~al.} 2021, The Astronomical Journal, 161, 122

\bibitem[{Hamdani {et~al.}(2006)Hamdani, Arnold, Foellmi, Berthier, Billeres, Briot, Fran{\c{c}}ois, Riaud, \& Schneider}]{hamdani2006biomarkers}
Hamdani, S., Arnold, L., Foellmi, C., {et~al.} 2006, Astronomy \& Astrophysics, 460, 617

\bibitem[{Hansen \& Hovenier(1974)}]{hansen1974interpretation}
Hansen, J.~E., \& Hovenier, J. 1974, Journal of Atmospheric Sciences, 31, 1137

\bibitem[{Hearty {et~al.}(2009)Hearty, Song, Kim, \& Tinetti}]{Hearty_2009}
Hearty, T., Song, I., Kim, S., \& Tinetti, G. 2009, The Astrophysical Journal, 693, 1763–1774

\bibitem[{Helling(2019)}]{helling2019exoplanet}
Helling, C. 2019, Annual Review of Earth and Planetary Sciences, 47, 583

\bibitem[{Kaltenegger {et~al.}(2007)Kaltenegger, Traub, \& Jucks}]{kaltenegger2007spectral}
Kaltenegger, L., Traub, W.~A., \& Jucks, K.~W. 2007, The Astrophysical Journal, 658, 598

\bibitem[{Kasting {et~al.}(1993)Kasting, Whitmire, \& Reynolds}]{kasting1993habitable}
Kasting, J.~F., Whitmire, D.~P., \& Reynolds, R.~T. 1993, Icarus, 101, 108

\bibitem[{Kawashima \& Rugheimer(2019)}]{kawashima2019theoretical}
Kawashima, Y., \& Rugheimer, S. 2019, The Astronomical Journal, 157, 213

\bibitem[{Keller-Rudek {et~al.}(2013)Keller-Rudek, Moortgat, Sander, \& S{\"o}rensen}]{keller2013mpi}
Keller-Rudek, H., Moortgat, G.~K., Sander, R., \& S{\"o}rensen, R. 2013, Earth System Science Data, 5, 365

\bibitem[{King {et~al.}(2013)King, Platnick, Menzel, Ackerman, \& Hubanks}]{king2013spatial}
King, M.~D., Platnick, S., Menzel, W.~P., Ackerman, S.~A., \& Hubanks, P.~A. 2013, IEEE transactions on geoscience and remote sensing, 51, 3826

\bibitem[{Kitzmann {et~al.}(2010)Kitzmann, Patzer, von Paris, Godolt, Stracke, Gebauer, Grenfell, \& Rauer}]{kitzmann2010clouds}
Kitzmann, D., Patzer, A., von Paris, P., {et~al.} 2010, Astronomy \& Astrophysics, 511, A66

\bibitem[{Kitzmann {et~al.}(2011{\natexlab{a}})Kitzmann, Patzer, von Paris, Godolt, \& Rauer}]{Kitzmann_2011b}
Kitzmann, D., Patzer, A. B.~C., von Paris, P., Godolt, M., \& Rauer, H. 2011{\natexlab{a}}, Astronomy \& Astrophysics, 534, A63

\bibitem[{Kitzmann {et~al.}(2011{\natexlab{b}})Kitzmann, Patzer, von Paris, Godolt, \& Rauer}]{Kitzmann_2011a}
---. 2011{\natexlab{b}}, Astronomy \& Astrophysics, 531, A62

\bibitem[{Kofman {et~al.}(2024)Kofman, Villanueva, Fauchez, Mandell, Johnson, Payne, Latouf, \& Kelkar}]{Kofman_2024}
Kofman, V., Villanueva, G.~L., Fauchez, T.~J., {et~al.} 2024, The Planetary Science Journal, 5, 197

\bibitem[{Kokaly {et~al.}(2017)Kokaly, Clark, Swayze, Livo, Hoefen, Pearson, Wise, Benzel, Lowers, Driscoll, {et~al.}}]{kokaly2017usgs}
Kokaly, R., Clark, R., Swayze, G., {et~al.} 2017, United States Geological Survey (USGS): Reston, VA, USA, 61

\bibitem[{Kopparapu {et~al.}(2021)Kopparapu, Arney, Haqq-Misra, Lustig-Yaeger, \& Villanueva}]{Kopparapu_2021_no2}
Kopparapu, R., Arney, G., Haqq-Misra, J., Lustig-Yaeger, J., \& Villanueva, G. 2021, The Astrophysical Journal, 908, 164

\bibitem[{Kreidberg {et~al.}(2014)Kreidberg, Bean, D{\'e}sert, Benneke, Deming, Stevenson, Seager, Berta-Thompson, Seifahrt, \& Homeier}]{kreidberg2014clouds}
Kreidberg, L., Bean, J.~L., D{\'e}sert, J.-M., {et~al.} 2014, Nature, 505, 69

\bibitem[{Line {et~al.}(2013)Line, Knutson, Deming, WILkINS, \& Desert}]{line2013near}
Line, M.~R., Knutson, H., Deming, D., WILkINS, A., \& Desert, J.-M. 2013, The Astrophysical Journal, 778, 183

\bibitem[{Loeb {et~al.}(2018)Loeb, Doelling, Wang, Su, Nguyen, Corbett, Liang, Mitrescu, Rose, \& Kato}]{loeb2018clouds}
Loeb, N.~G., Doelling, D.~R., Wang, H., {et~al.} 2018, Journal of climate, 31, 895

\bibitem[{{LUVOIR Team}(2019)}]{luvoir2019luvoir}
{LUVOIR Team}. 2019, arXiv preprint arXiv:1912.06219

\bibitem[{Marley {et~al.}(2013)Marley, Ackerman, Cuzzi, \& Kitzmann}]{marley2013clouds}
Marley, M.~S., Ackerman, A.~S., Cuzzi, J.~N., \& Kitzmann, D. 2013, Comparative climatology of terrestrial planets, 1, 367

\bibitem[{Marley {et~al.}(1999)Marley, Gelino, Stephens, Lunine, \& Freedman}]{marley1999reflected}
Marley, M.~S., Gelino, C., Stephens, D., Lunine, J.~I., \& Freedman, R. 1999, The Astrophysical Journal, 513, 879

\bibitem[{May {et~al.}(2021)May, Taylor, Komacek, Line, \& Parmentier}]{May_2021}
May, E.~M., Taylor, J., Komacek, T.~D., Line, M.~R., \& Parmentier, V. 2021, The Astrophysical Journal Letters, 911, L30

\bibitem[{Montanes-Rodriguez {et~al.}(2006)Montanes-Rodriguez, Pall{\'e}, Goode, \& Mart{\'\i}n-Torres}]{montanes2006vegetation}
Montanes-Rodriguez, P., Pall{\'e}, E., Goode, P., \& Mart{\'\i}n-Torres, F. 2006, The Astrophysical Journal, 651, 544

\bibitem[{Montmessin {et~al.}(2007)Montmessin, Gondet, Bibring, Langevin, Drossart, Forget, \& Fouchet}]{montmessin2007hyperspectral}
Montmessin, F., Gondet, B., Bibring, J.-P., {et~al.} 2007, Journal of Geophysical Research: Planets, 112

\bibitem[{{National Academies of Sciences, Engineering, and Medicine}(2023)}]{decadal_survey}
{National Academies of Sciences, Engineering, and Medicine}. 2023, {Pathways to Discovery in Astronomy and Astrophysics for the 2020s} (Washington, DC: The National Academies Press)

\bibitem[{Powell {et~al.}(2019)Powell, Louden, Kreidberg, Zhang, Gao, \& Parmentier}]{Powell_2019}
Powell, D., Louden, T., Kreidberg, L., {et~al.} 2019, The Astrophysical Journal, 887, 170

\bibitem[{{Roberge} {et~al.}(2012){Roberge}, {Chen}, {Millan-Gabet}, {Weinberger}, {Hinz}, {Stapelfeldt}, {Absil}, {Kuchner}, \& {Bryden}}]{2012_Roberge_Dust}
{Roberge}, A., {Chen}, C.~H., {Millan-Gabet}, R., {et~al.} 2012, \pasp, 124, 799

\bibitem[{Rossow \& Lacis(1990)}]{rossow1990global}
Rossow, W.~B., \& Lacis, A.~A. 1990, Journal of Climate, 3, 1204

\bibitem[{Rothman {et~al.}(2010)Rothman, Gordon, Barber, Dothe, Gamache, Goldman, Perevalov, Tashkun, \& Tennyson}]{rothman2010hitemp}
Rothman, L.~S., Gordon, I., Barber, R., {et~al.} 2010, Journal of Quantitative Spectroscopy and Radiative Transfer, 111, 2139

\bibitem[{S{\'a}nchez-Lavega {et~al.}(2004)S{\'a}nchez-Lavega, P{\'e}rez-Hoyos, \& Hueso}]{sanchez2004clouds}
S{\'a}nchez-Lavega, A., P{\'e}rez-Hoyos, S., \& Hueso, R. 2004, American Journal of Physics, 72, 767

\bibitem[{Saxena(2022)}]{Saxena_2022}
Saxena, P. 2022, The Astrophysical Journal Letters, 934, L32

\bibitem[{Saxena {et~al.}(2021)Saxena, Villanueva, Zimmerman, Mandell, \& Smith}]{Saxena_2021}
Saxena, P., Villanueva, G.~L., Zimmerman, N.~T., Mandell, A.~M., \& Smith, A. J. R.~W. 2021, The Astronomical Journal, 162, 30

\bibitem[{Segura {et~al.}(2005)Segura, Kasting, Meadows, Cohen, Scalo, Crisp, Butler, \& Tinetti}]{segura2005biosignatures}
Segura, A., Kasting, J.~F., Meadows, V., {et~al.} 2005, Astrobiology, 5, 706

\bibitem[{Segura {et~al.}(2003)Segura, Krelove, Kasting, Sommerlatt, Meadows, Crisp, Cohen, \& Mlawer}]{segura2003ozone}
Segura, A., Krelove, K., Kasting, J.~F., {et~al.} 2003, Astrobiology, 3, 689

\bibitem[{Sing {et~al.}(2015)Sing, Wakeford, Showman, Nikolov, Fortney, Burrows, Ballester, Deming, Aigrain, D{\'e}sert, {et~al.}}]{sing2015hst}
Sing, D.~K., Wakeford, H.~R., Showman, A.~P., {et~al.} 2015, Monthly Notices of the Royal Astronomical Society, 446, 2428

\bibitem[{Sneep \& Ubachs(2005)}]{sneep2005direct}
Sneep, M., \& Ubachs, W. 2005, Journal of Quantitative Spectroscopy and Radiative Transfer, 92, 293

\bibitem[{Stamnes {et~al.}(1988)Stamnes, Tsay, Wiscombe, \& Jayaweera}]{stamnes1988numerically}
Stamnes, K., Tsay, S.-C., Wiscombe, W., \& Jayaweera, K. 1988, Applied optics, 27, 2502

\bibitem[{Stamnes {et~al.}(2000)Stamnes, Tsay, Wiscombe, \& Laszlo}]{stamnes2000disort}
Stamnes, K., Tsay, S.-C., Wiscombe, W., \& Laszlo, I. 2000

\bibitem[{Stark {et~al.}(2024)Stark, Latouf, Mandell, \& Young}]{10.1117/1.JATIS.10.1.014005}
Stark, C.~C., Latouf, N., Mandell, A.~M., \& Young, A. 2024, Journal of Astronomical Telescopes, Instruments, and Systems, 10, 014005

\bibitem[{Stowe {et~al.}(1991)Stowe, McClain, Carey, Pellegrino, Gutman, Davis, Long, \& Hart}]{stowe1991global}
Stowe, L., McClain, E., Carey, R., {et~al.} 1991, Advances in Space Research, 11, 51

\bibitem[{Sudarsky {et~al.}(2003)Sudarsky, Burrows, \& Hubeny}]{Sudarsky_2003}
Sudarsky, D., Burrows, A., \& Hubeny, I. 2003, The Astrophysical Journal, 588, 1121–1148

\bibitem[{Tinetti {et~al.}(2006{\natexlab{a}})Tinetti, Rashby, \& Yung}]{tinetti2006detectability}
Tinetti, G., Rashby, S., \& Yung, Y.~L. 2006{\natexlab{a}}, The Astrophysical Journal, 644, L129

\bibitem[{Tinetti {et~al.}(2006{\natexlab{b}})Tinetti, Rashby, \& Yung}]{tinetti2006b}
---. 2006{\natexlab{b}}, The Astrophysical Journal, 644, L129

\bibitem[{Vasquez {et~al.}(2013)Vasquez, Schreier, Garc{\'\i}a, Kitzmann, Patzer, Rauer, \& Trautmann}]{vasquez2013infrared}
Vasquez, M., Schreier, F., Garc{\'\i}a, S.~G., {et~al.} 2013, Astronomy \& Astrophysics, 557, A46

\bibitem[{Villanueva {et~al.}(2015)Villanueva, Mumma, Novak, K{\"a}ufl, Hartogh, Encrenaz, Tokunaga, Khayat, \& Smith}]{villanueva2015strong}
Villanueva, G., Mumma, M., Novak, R., {et~al.} 2015, Science, 348, 218

\bibitem[{{Villanueva} {et~al.}(2022){Villanueva}, {Liuzzi}, {Faggi}, {Protopapa}, {Kofman}, {Fauchez}, {Stone}, \& {Mandell}}]{psg_handbook}
{Villanueva}, G.~L., {Liuzzi}, G., {Faggi}, S., {et~al.} 2022, {Fundamentals of the Planetary Spectrum Generator}

\bibitem[{Villanueva {et~al.}(2018)Villanueva, Smith, Protopapa, Faggi, \& Mandell}]{villanueva2018planetary}
Villanueva, G.~L., Smith, M.~D., Protopapa, S., Faggi, S., \& Mandell, A.~M. 2018, Journal of Quantitative Spectroscopy and Radiative Transfer, 217, 86

\bibitem[{Wang {et~al.}(2017)Wang, Mawet, Ruane, Delorme, Klimovich, \& Hu}]{wang2017baseline}
Wang, J., Mawet, D., Ruane, G., {et~al.} 2017, in Techniques and Instrumentation for Detection of Exoplanets VIII, Vol. 10400, SPIE, 253--261

\bibitem[{Wolff {et~al.}(2009)Wolff, Smith, Clancy, Arvidson, Kahre, Seelos~Iv, Murchie, \& Savij{\"a}rvi}]{wolff2009wavelength}
Wolff, M., Smith, M., Clancy, R., {et~al.} 2009, Journal of Geophysical Research: Planets, 114

\bibitem[{Wu {et~al.}(2011)Wu, Hu, McCormick, \& Yan}]{wu2011global}
Wu, D., Hu, Y., McCormick, M.~P., \& Yan, F. 2011, International Journal of Remote Sensing, 32, 1269

\bibitem[{Wylie {et~al.}(2005)Wylie, Jackson, Menzel, \& Bates}]{wylie2005trends}
Wylie, D., Jackson, D.~L., Menzel, W.~P., \& Bates, J.~J. 2005, Journal of climate, 18, 3021

\bibitem[{Young {et~al.}(2018)Young, Knapp, Inamdar, Hankins, \& Rossow}]{isccp_dataset}
Young, A.~H., Knapp, K.~R., Inamdar, A., Hankins, W., \& Rossow, W.~B. 2018, Earth System Science Data, 10, 583

\end{thebibliography}
\bibliographystyle{aasjournal}



\end{document}